\documentclass[a4paper,11pt,oneside]{article}
\usepackage{float}
\usepackage{graphicx}
\usepackage{amsfonts,amssymb,amsmath,amsthm}
\usepackage{enumerate}
\usepackage{subcaption}
\usepackage{placeins}
\usepackage[table]{xcolor}
\usepackage{capt-of}
\usepackage{booktabs}
\usepackage{caption}
\usepackage{tabularx}
\usepackage{array}
\usepackage{enumitem}
\usepackage{tikz}
\usepackage{pgfplots}
\usepackage[numbers,sort&compress]{natbib}
\usepackage[bookmarks,colorlinks,pdfstartview=FitH]{hyperref}
\usepackage[left=1in, right=1in, top=1in, bottom=1in]{geometry}

\hypersetup{
  citecolor=blue,
  filecolor=black,
  linkcolor=blue,
  urlcolor=black
}

\theoremstyle{plain}

\theoremstyle{definition}

\theoremstyle{remark}
\newtheorem{remark}{Remark}[section]

\title{\large\textbf{
Resonance Induced Chaos of Charged Particles Around a Magnetized, Periodically Time-Conformal Reissner–Nordström Black Hole with Quintessence}}

\author{\footnotesize
Israr Ali Khan\textsuperscript{1,$\ast$}\\[2mm]
\scriptsize\textsuperscript{1}Institute of Numerical Sciences, Kohat University of Science and Technology, Kohat, KPK, Pakistan\\[2mm]
\scriptsize\textsuperscript{*}Correspondence: israrali@kust.edu.pk
}

\date{}

\begin{document}

\maketitle

\begin{abstract}
\noindent
Astrophysical black holes are never strictly static: accretion, mergers, and gravitational radiation drive a slow evolution of their geometry. We model this evolution through a periodic time-conformal deformation, $\Omega(t)=1+\epsilon\sin(\omega_d t)$, of a Reissner--Nordstr\"om black hole surrounded by quintessence, and study the resulting onset of chaos for charged test particles. The deformation converts the integrable static spacetime into a periodically driven nonlinear Hamiltonian system, so that the transition from regular to chaotic motion can be related to the breakdown of an approximate Noether symmetry. The resulting two-degree-of-freedom Hamiltonian system is solved using an adaptive high-order Runge--Kutta integrator, while the dynamical behaviour is analysed through Poincar\'e sections, bifurcation diagrams, the largest Lyapunov exponent, and the Smaller Alignment Index (SALI). Our results show that the onset of chaos is controlled primarily by nonlinear resonance rather than by the magnitude of any individual parameter. Chaos grows with the conformal amplitude $\epsilon$ through resonance overlap, while the magnetic field $B$ acts non-monotonically: it produces chaos near the field strength at which the radial frequency matches the drive, and suppresses it at stronger fields by detuning. The small-divisor structure of the approximate Noether symmetry predicts the magnetic threshold without free parameters ($B_c\simeq0.011$ against the measured $0.012$), and reproduces the conformal threshold once a single coupling constant is fixed. In contrast, the quintessence parameter mainly facilitates particle escape by reducing the confining potential barrier, providing a distinct non-resonant route to the breakdown of regular motion.
\end{abstract}

\section{Introduction}
\label{sec:intro}

The dynamics of test particles in the vicinity of black holes provides a natural framework for investigating the interplay between gravity, electromagnetism, and nonlinear dynamics. Geodesic motion in the Schwarzschild, Reissner--Nordstr\"om, and Kerr spacetimes is fully integrable because these geometries admit a complete set of conserved quantities, the last of which follows in the Kerr case from the hidden symmetry encoded in the Carter constant \citep{Carter1968}. This integrability, however, is highly sensitive to external perturbations. The introduction of magnetic fields, surrounding matter distributions, or explicit time dependence in the spacetime metric generally breaks one or more conservation laws, giving rise to non-integrable behaviour and chaotic particle motion \citep{Kiselev2003,Dalui2019,TakahashiKoyama2009,KopacekEtal2010}. Understanding the transition from regular to chaotic motion is therefore important not only from the perspective of relativistic dynamics but also for astrophysical phenomena, since chaotic scattering can energize charged particles and thereby contribute to particle acceleration and jet formation in black-hole magnetospheres \citep{StuchlikKolos2016,Tursunov2020}.

Two developments of the past two decades are directly relevant here. The first is the solution proposed by Kiselev, which describes a black hole embedded in a quintessential dark-energy background through a simple yet analytically tractable spacetime geometry \citep{Kiselev2003}. Although the associated stress--energy tensor does not correspond to that of a fundamental quintessence field \citep{Visser2020}, the model has become a widely used framework for investigating the influence of dark energy on black-hole spacetimes \citep{Singh2023,HamilLutfuoglu2024}. The second development arose from the realization that the growth rate of chaos near an event horizon is bounded by the surface gravity, a result established in the holographic framework in \citep{MSS2016} and later recovered directly from the near-horizon dynamics of test particles \citep{HashimotoTanahashi2017}. It has since motivated many studies of chaotic particle motion in the near-horizon region, including studies demonstrating that the presence of an event horizon alone can render an otherwise integrable system chaotic \citep{Dalui2019}, together with analyses of the conditions under which the bound is saturated or violated for charged particles around charged black holes \citep{ZhaoLiLu2018,KanGwak2022}. Meanwhile, numerical studies of charged particles around magnetized black holes have identified magnetic-field-induced resonances and the breaking of spatial symmetries as primary mechanisms responsible for the onset of chaos and for the structure of the associated escape zones \citep{SunWangLiuWu2021,Shaymatov2021,KopacekKaras2018,Panis2019,Yang2023Chaos}.

A complementary line of research has focused specifically on how explicit time dependence in the background geometry reshapes particle motion. The idea of promoting a static black-hole metric to a slowly evolving one through a time-conformal factor, selected by the requirement of an approximate Noether symmetry, was introduced by Jawad \emph{et al.} for the Schwarzschild spacetime \citep{Jawad2016}. In this construction the conformal factor rescales the energy content of the spacetime and encodes, at the level of a weakly broken symmetry, the slow drain of energy associated with processes such as gravitational radiation and Hawking emission. It was subsequently applied to the AdS--Schwarzschild and quantum-corrected Schwarzschild geometries \citep{Khan2019PDU}, and to the quintessential Schwarzschild spacetime \citep{Khan2020IJMPD}. More recently, the same construction has been carried over to charged backgrounds, first to the AdS Reissner--Nordstr\"om black hole \citep{Khan2024MPLA} and then to the perturbed Reissner--Nordstr\"om--AdS geometry surrounded by quintessential dark energy \citep{Khan2025IJGMMP}. Taken together, these works established the effective potential, escape velocity, and orbital stability of the perturbed spacetimes, and in several cases used the Lyapunov exponent as a first local indicator of orbital instability. What they share, however, is a treatment of the time-conformal perturbation mainly as a device for rescaling conserved quantities and probing local stability, rather than as a periodic driving capable of reorganizing the global phase-space structure of the motion. The transition from regular to genuinely chaotic dynamics, and how it relates to the order at which the approximate Noether symmetry breaks down, has therefore remained largely unexplored.

Outside this line of work, the chaos literature reviewed above treats the background spacetime as strictly static. In realistic astrophysical environments, however, black holes are rarely perfectly stationary. Their masses and surrounding geometries evolve continually through accretion, mergers, and the emission of gravitational waves, producing deviations from an exactly static spacetime. A simple yet physically motivated way to model such slow evolution, while preserving the analytical advantages of a spherically symmetric spacetime, is to introduce a slowly varying time-dependent conformal factor. This approach has been developed in a series of studies based on approximate Noether symmetry, where the loss of exact energy conservation is described through a weakly broken symmetry whose breaking encodes the energy carried away by gravitational radiation \citep{Khan2022T,Khan2024MPLA}. When the conformal factor is chosen to oscillate periodically in time, the particle dynamics is transformed from an autonomous system into a periodically driven nonlinear oscillator. Such systems provide the classical framework for understanding the transition from regular to chaotic motion through nonlinear resonance overlap, as described by the Chirikov resonance-overlap criterion \citep{Chirikov1979} and the Kolmogorov--Arnold--Moser (KAM) theorem \citep{Arnold1963}. This motivates the central question addressed in the present work: \emph{is the onset of chaos in a time-conformal black-hole spacetime governed by the same resonance structure that underlies the breakdown of approximate Noether symmetry?} If such a connection exists, the perturbative order at which the approximate symmetry breaks down would provide an analytical prediction of the chaos threshold, endowing the approximate Noether symmetry with not only a descriptive role but also a quantitative measure of dynamical stability.

In this work, we investigate this correspondence by considering the motion of a charged test particle confined to the equatorial plane of a Reissner--Nordstr\"om--quintessence black hole \citep{Kiselev2003}, immersed in a weak external magnetic field described by the standard Wald test-field construction \citep{Wald1974}, and subjected to a periodic time-conformal deformation of the form $\Omega(t)=1+\epsilon\sin(\omega_d t)$. The resulting equations of motion are formulated as a periodically driven two-degree-of-freedom Hamiltonian system and integrated numerically using a high-order adaptive Runge--Kutta method. The transition from regular to chaotic motion is characterized through stroboscopic Poincar\'e sections, bifurcation diagrams, the largest Lyapunov exponent computed using the algorithm of Benettin \emph{et al.}~\citep{Benettin1980}, and the Smaller Alignment Index (SALI) introduced by Skokos \citep{Skokos2001}. The numerically identified chaos thresholds are then compared with the analytical resonance conditions and the small-divisor structure underlying approximate Noether symmetry. We find that both the conformal amplitude $\epsilon$ and the magnetic field $B$ act through nonlinear resonance: in $\epsilon$, chaos sets in through the overlap of neighbouring resonances, while the magnetic threshold closely follows the crossing condition $\Omega_{\rm orb}=\omega_d$. On the other hand, the quintessence parameter $\alpha$ mainly affects the dynamics by lowering the potential barrier and thereby facilitating orbital escape. This is consistent with the escape behaviour reported earlier for the corresponding static spacetime \citep{Khan2021R}.

The rest of the paper is organized as follows. Section~\ref{sec:background} introduces the time-conformal Reissner--Nordstr\"om black hole surrounded by quintessence and describes its horizon structure together with the external Wald magnetic field. The Lagrangian and Hamiltonian descriptions along with the exact conservation laws and the approximate energy invariant, the equations of motion and the effective potential are covered in Section~\ref{sec:hamiltonian}. The conformal driving functions, the numerical methods and chaos indicators used are described in Section~\ref{sec:numerics}. Section~\ref{sec:results} presents the numerical results, including the static limit, the effects of the conformal amplitude $\epsilon$, the quintessence parameter $\alpha$, and the magnetic field $B$, together with the global Lyapunov and SALI maps and the transient dynamics under monotonic driving. Section~\ref{sec:correlation} examines the connection between the numerically identified chaos thresholds and the order-by-order breakdown of approximate Noether symmetry through their common small-divisor resonance structure. Finally, Section~\ref{sec:conclusion} summarizes the main results and discusses their broader astrophysical implications.

\bigskip

\section{Spacetime and Electromagnetic Background}
\label{sec:background}

We consider a static, spherically symmetric Reissner--Nordstr\"om black hole surrounded by a quintessential dark-energy field, whose geometry is subsequently deformed by a slowly varying time-dependent conformal factor. Time-conformal metrics provide a simple yet physically motivated framework for describing black holes embedded in an evolving cosmological environment while preserving the analytical tractability of a spherically symmetric spacetime \citep{Jawad2016,Khan2025IJGMMP}. The line element is given by
\begin{equation}
ds^{2} = \Omega^{2}(t)\left[-f(r)\,dt^{2} + \frac{dr^{2}}{f(r)}
+ r^{2}d\theta^{2} + r^{2}\sin^{2}\theta\, d\phi^{2}\right],
\label{eq:metric}
\end{equation}
where $\Omega(t)$ is a dimensionless, positive conformal factor whose deviation from unity is controlled by the small parameter $\epsilon$ (introduced explicitly in Sec.~\ref{sec:numerics}). The assumption of slow temporal evolution is required only for the adiabatic approximate-Noether-symmetry analysis presented in Sec.~\ref{sec:hamiltonian}; in the numerical study, the driving frequency is allowed to span both resonant and non-resonant regimes in order to investigate the transition to chaos. The metric function is
\begin{equation}
f(r) = 1 - \frac{2M}{r} + \frac{Q^{2}}{r^{2}}
- \frac{\alpha}{r^{3\omega_{q}+1}},
\label{eq:f}
\end{equation}
where the quintessential correction term follows the construction derived by Kiselev \citep{Kiselev2003}. Here, $M$ and $Q$ denote the black-hole mass and electric charge, $\alpha$ characterizes the intensity of the surrounding quintessence field, and $\omega_q$ is its equation-of-state parameter. Throughout this work we consider the quintessence regime $-1<\omega_q<-1/3$, for which the field exhibits dark-energy-like behaviour with negative pressure. In the limit $\alpha\rightarrow0$, Eq.~\eqref{eq:f} reduces to the standard Reissner--Nordstr\"om geometry, and setting $Q\rightarrow0$ in addition recovers the Schwarzschild solution. When $\Omega(t)\equiv1$, the metric coincides with the static background investigated in our previous study of neutral and charged particle dynamics \citep{Khan2021R}. The present model extends that framework by promoting the conformal factor to an explicit function of time, following the approach adopted in our earlier studies of time-conformal black-hole thermodynamics and orbital dynamics \citep{Khan2022T,Khan2025IJGMMP}.

The horizon structure is determined entirely by the static metric function $f(r)$, since the conformal factor $\Omega^{2}(t)>0$ rescales the metric without changing the locations of the null hypersurfaces defined by $f(r)=0$. Depending on the values of $M$, $Q$, and $\alpha$, Eq.~\eqref{eq:f} admits up to three positive roots, denoted by $r_{-}<r_{+}<r_q$, corresponding to the inner Cauchy horizon, the event horizon, and the outer quintessential horizon, respectively. The presence of an outer horizon results from the large-$r$ asymptotics of the quintessence term: for $-1<\omega_q<-1/3$ its magnitude grows with $r$, so that $f(r)$ eventually crosses zero beyond the event horizon. Consequently, $r_q$ acts as a cosmological-type boundary analogous to the de~Sitter horizon. Throughout this work, the parameter $\alpha$ is restricted to values for which all three horizons remain distinct, and all particle trajectories are initialized within the physically accessible region $r_{+}<r<r_q$, thereby preserving the causal structure of the spacetime during the conformal evolution.

An external asymptotically uniform magnetic field is incorporated using the test-field construction proposed by Wald
\citep{Wald1974}. On a vacuum background this construction yields an
exact solution of the source-free Maxwell equations; on the present
non-vacuum (charged, quintessence-supported) geometry it remains a
consistent weak-field approximation, valid to leading order in $B$ in
the exterior region of interest, and we adopt it in that spirit
without back-reaction on the metric. We further note that, since the
Maxwell equations are conformally invariant in four dimensions, the
potential \eqref{eq:Aphi} is unaffected by the time-conformal factor
$\Omega(t)$. The Coulomb potential $A_t = Q/r$ sourced by the
black hole's own charge is deliberately switched off in the particle
sector: we restrict attention to the purely magnetic coupling, so that
the parameter $Q$ enters the dynamics only geometrically through
$f(r)$. Including the electrostatic interaction is a straightforward
extension left for future work. In the
present spherically symmetric setting this construction yields a single
non-vanishing component of the vector potential,
\begin{align}
A_{\phi} &= \tfrac{1}{2}B\,r^{2}\sin^{2}\theta, \nonumber \\
A_{t} &= A_{\theta} = A_{r} = 0,
\label{eq:Aphi}
\end{align}
where $B$ denotes the asymptotic strength of the magnetic field as
measured by a static observer at infinity. Because the magnetic field is
treated as a test field, its presence does not alter the metric
functions $\Omega(t)$ or $f(r)$; it enters the particle dynamics only
through the minimal coupling term $qA_{\mu}\dot{x}^{\mu}$ in the
Lagrangian, exactly as in our earlier treatment of charged particle
motion in the static case \citep{Khan2021R}; see also \citep{Wang2023Quintessence}.

\begin{figure}[t]
\centering
\includegraphics[width=\textwidth]{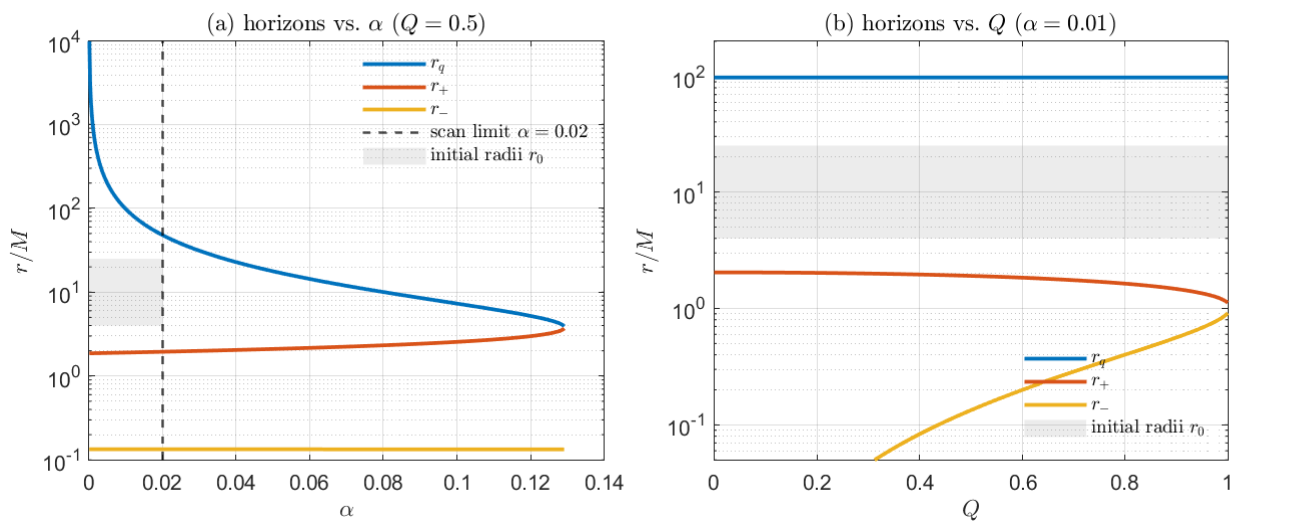}
\caption{\footnotesize \textit{\textbf{Horizon structure.} Locations of the roots of $f(r)=0$ as functions of (a) the quintessence parameter $\alpha$ for $Q=0.5$ and (b) the black-hole charge $Q$ for $\alpha=0.01$, with $\omega_q=-2/3$. The dashed line in panel (a) indicates the maximum value $\alpha=10^{-3}$ considered in the numerical study, which remains well below the horizon-merger limit ($\alpha\approx0.13$). The shaded region denotes the range of initial radii $r_0$ used in the numerical simulations, confirming that all trajectories are initialized within the physically accessible region bounded by the event horizon $r_+$ and the quintessential horizon $r_q$.}}
\label{fig:horizon}
\end{figure}

Since both the metric \eqref{eq:metric} and the vector potential \eqref{eq:Aphi} are symmetric under reflection about the equatorial plane ($\theta=\pi/2$), any geodesic or charged-particle trajectory initialized on this plane with vanishing polar velocity remains confined to it throughout the motion. We therefore restrict the analysis to equatorial motion, $\theta=\pi/2$ and $\dot{\theta}=0$, without loss of generality. This simplification was also adopted in our earlier studies \citep{Khan2020IJMPD} and is widely used in investigations of relativistic chaos, as it reduces the system to the minimum number of degrees of freedom required for Poincar\'e-section analysis while preserving the essential nonlinear coupling between the radial motion, angular momentum, and the magnetic field. The horizon structure corresponding to the parameter ranges considered in the numerical study is shown in Fig.~\ref{fig:horizon}.

\section{Lagrangian, Hamiltonian, and Exact/Approximate Conservation Laws}
\label{sec:hamiltonian}

Restricting the motion to the equatorial plane ($\theta=\pi/2$), as justified in Sec.~\ref{sec:background}, the dynamics of a charged test particle with unit mass and electric charge $q$ in the spacetime \eqref{eq:metric} is described by the minimally coupled Lagrangian
\begin{equation}
\mathcal{L} = \frac{1}{2}\Omega^{2}(t)\left[-f(r)\dot t^{2}
+ \frac{\dot r^{2}}{f(r)} + r^{2}\dot\phi^{2}\right] + qA_{\phi}\dot\phi .
\label{eq:lagrangian}
\end{equation}
Here and throughout the paper, an overdot denotes differentiation with respect to the particle proper time $\tau$, whereas the comma notation, $\Omega_{,t}\equiv d\Omega/dt$, denotes differentiation with respect to the coordinate time $t$. The two derivatives are related through $d\Omega/d\tau=\Omega_{,t}\dot t$ and should therefore not be confused. Since the coordinate $\phi$ is cyclic, its conjugate momentum is exactly conserved,
\begin{equation}
P_{\phi} \equiv L = \Omega^{2}(t)\,r^{2}\dot\phi + qA_{\phi} =
\text{const.},
\label{eq:Pphi}
\end{equation}
and is identified with the particle's conserved orbital angular momentum. This conservation law remains unaffected by the conformal deformation because $\Omega(t)$ enters the Lagrangian only as an overall multiplicative factor and is independent of $\phi$. Together with the mass-shell constraint introduced below, it reduces the dynamics to an effectively two-degree-of-freedom system, whose reduced $(r,P_r)$ plane is used for the Poincar\'e-section analysis presented in Sec.~\ref{sec:numerics}.

The situation is different for the time coordinate. Because the conformal factor depends explicitly on $t$, the Lagrangian \eqref{eq:lagrangian} is no longer invariant under time translations, and the particle energy therefore ceases to be an exact constant of motion. Nevertheless, following the approximate Noether symmetry formalism developed in~\citep{Jawad2016}, based on the perturbative framework introduced by Baikov \emph{et al.}~\citep{Baikov1988}, an adiabatic energy invariant $\tilde{E}(t)$ can still be constructed order by order in the conformal strength $\epsilon$. Since this construction has been presented in detail elsewhere, it is not repeated here. Instead, Sec.~\ref{sec:correlation} examines how the perturbative breakdown of this approximate invariant correlates with the numerically observed onset of chaos.

For the numerical analysis, it is more convenient to adopt the Hamiltonian formulation. Using the mass-shell condition
$g^{\mu\nu}(P_\mu-qA_\mu)(P_\nu-qA_\nu)=-1$,
which is satisfied by the kinetic four-momentum of a timelike charged particle \citep{MTW1973}, the Hamiltonian takes the form
\begin{equation}
H = \frac{1}{2\Omega^{2}(t)}\left[-\frac{P_{t}^{2}}{f(r)} + f(r)P_{r}^{2}
+ \frac{(P_{\phi}-qA_{\phi})^{2}}{r^{2}}\right] = -\frac{1}{2},
\label{eq:Hamiltonian}
\end{equation}
and Hamilton's equations then give a closed system of six
first-order equations,
\begin{subequations}
\label{eq:eom}
\begin{align}
\dot t &= -\frac{P_{t}}{\Omega^{2}f}, \label{eq:eom_t}\\
\dot r &= \frac{fP_{r}}{\Omega^{2}}, \label{eq:eom_r}\\
\dot\phi &= \frac{P_{\phi}-qA_{\phi}}{\Omega^{2}r^{2}}, \label{eq:eom_phi}\\
\dot P_{t} &= -\frac{\Omega_{,t}}{\Omega}, \label{eq:eom_Pt}\\
\dot P_{r} &= -\frac{1}{2\Omega^{2}}\!\left[\frac{f'}{f^{2}}P_{t}^{2}
+ f'P_{r}^{2}\right] + \frac{qA_\phi'(P_\phi-qA_\phi)}{\Omega^{2}r^{2}}
+ \frac{(P_\phi-qA_\phi)^{2}}{\Omega^{2}r^{3}}, \label{eq:eom_Pr}\\
\dot P_{\phi} &= 0. \label{eq:eom_Pphi}
\end{align}
\end{subequations}
Fixing the conserved $P_\phi$ by its initial value leaves five equations, and since the cyclic variable $\phi$ decouples from the rest of the flow, the dynamics relevant for chaos lives in the four variables $(t,r,P_t,P_r)$. These equations are the basis of all numerical integrations below.

Equation~\eqref{eq:eom_Pt} plays a central role in the present analysis. It follows directly from Hamilton's equation $\dot P_t=-\partial H/\partial t$ together with the mass-shell constraint $H=-1/2$, and is therefore exact rather than perturbative. Since the conserved quantity associated with time translations is $E=-P_t$, Eq.~\eqref{eq:eom_Pt} may be written equivalently as
\begin{equation}
\dot E=\frac{\Omega_{,t}}{\Omega},
\label{eq:Edot}
\end{equation}
which explicitly describes the exchange of energy between the particle and the time-dependent spacetime. In the static limit, where $\Omega=\mathrm{constant}$, the right-hand side vanishes and the particle energy is recovered as an exact constant of motion. The explicit time dependence of the conformal factor is therefore what breaks integrability and drives the dynamics studied below.
\begin{remark}
Although Eq.~\eqref{eq:eom_Pt} is the form valid on the constraint
surface $H=-1/2$, the numerical implementation integrates the
off-shell Hamiltonian form $\dot P_t = 2H\,\Omega_{,t}/\Omega$, which
reduces to Eq.~\eqref{eq:eom_Pt} on shell. This keeps the monitored
constraint violation $|H+1/2|$ a genuinely independent diagnostic of
the integration accuracy rather than a quantity partially enforced by
construction.
\end{remark}

\subsection{Effective potential}

\begin{figure}[H]
\centering
\includegraphics[width=\textwidth]{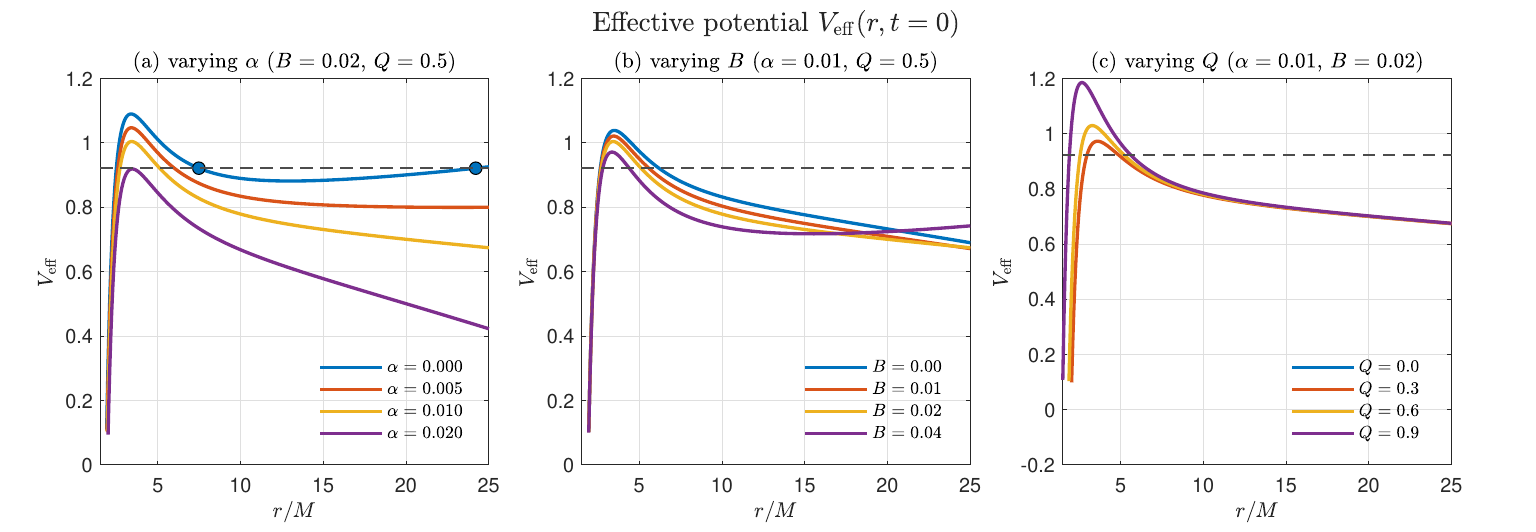}
\caption{\footnotesize \textit{\textbf{Effective potential.}
Effective potential $V_{\rm eff}(r,t)$ defined by Eq.~\eqref{eq:Veff} at $t=0$, corresponding to the static limit ($\Omega=1$), for $L=4.3$ and $E=0.96$ (horizontal dashed line). Panel (a) shows the effect of varying the quintessence parameter $\alpha$ ($B=0.02$, $Q=0.5$), panel (b) the magnetic field $B$ ($\alpha=0.01$, $Q=0.5$), and panel (c) the black-hole charge $Q$ ($\alpha=0.01$, $B=0.02$). Filled circles indicate radial turning points ($P_r=0$) that bound the allowed region of motion, while curves without markers do not support bound orbits at this energy. }}
\label{fig:Veff}
\end{figure}

Solving the mass-shell constraint \eqref{eq:Hamiltonian} for $P_t^2$
gives
\begin{align}
P_{t}^{2} &= V_{\rm eff}(r,t) + f^{2}(r)\,P_{r}^{2}, \nonumber \\
V_{\rm eff}(r,t) &= f(r)\left[\Omega^{2}(t) + \frac{(P_{\phi}-qA_{\phi})^{2}}{r^{2}}\right],
\label{eq:Veff}
\end{align}
which is the natural time-dependent extension of the static effective potential derived in \citep{Khan2021R}. Equation~\eqref{eq:Veff} is used to determine the physically admissible initial conditions for all numerical integrations. For prescribed values of $L$, $\alpha$, $B$, and the initial time $t_0$, the radial turning points are obtained from the condition $P_r=0$, giving $P_t^2=V_{\rm eff}(r,t_0)$. This relation is used to select the initial radius $r_0$ (Fig.~\ref{fig:Veff}) and to verify throughout the numerical evolution that the Hamiltonian constraint \eqref{eq:Hamiltonian} is satisfied to the accuracy reported in Sec.~\ref{sec:numerics}.

\section{Choice of Conformal Driving and Numerical Setup}
\label{sec:numerics}

To construct stroboscopic Poincar\'e sections, the conformal factor must be specified explicitly rather than treated as a generic slowly varying function. We therefore adopt the periodic form
\begin{equation}
\Omega(t) = 1 + \epsilon\sin(\omega_{d}t),
\label{eq:Omega}
\end{equation}
which transforms the charged-particle dynamics into the relativistic analogue of a periodically driven nonlinear oscillator, similar to the Duffing and H\'enon--Heiles systems that serve as canonical models of order--chaos transitions in nonlinear dynamics \citep{Henon1964,GuckenheimerHolmes1983}. Here, $\epsilon$ determines the perturbation strength and hence the degree of breaking of the approximate energy symmetry discussed in Sec.~\ref{sec:hamiltonian}, whereas $\omega_d$ denotes the driving frequency, whose ratio to the unperturbed orbital frequency governs the resonance structure. To examine the transient response to a smooth, non-periodic perturbation, we also consider the complementary secular driving profile
\begin{equation}
\Omega(t) = 1 + \epsilon\tanh(t/\tau_{s}),
\label{eq:Omega_transient}
\end{equation}
employed in the supplementary numerical study presented in Sec.~\ref{sec:transient}.

The equations of motion \eqref{eq:eom} are integrated using the adaptive eighth-order Runge--Kutta scheme DOP853 with embedded error control \citep{Hairer1993}. Although DOP853 is not symplectic, it provides sufficient accuracy to preserve the Hamiltonian constraint over the long integration times considered in this work. The constraint violation, $|H(t)+1/2|$, is monitored throughout every integration, and trajectories exceeding the tolerance $10^{-10}$ are recomputed using stricter error tolerances. Unless stated otherwise, the calculations are performed with $\omega_q=-2/3$, $Q=0.5$, $q=1$, $M=1$, and unit particle mass. The parameter ranges explored are $0\le\alpha\le10^{-3}$, $0\le B\le0.015$, $0\le\epsilon\le0.05$, while the driving frequency is varied over approximately one decade below and above the unperturbed radial epicyclic frequency. Except where the growing quintessence term erodes the outer barrier (Sec.~\ref{sec:alpha_results}), these ranges preserve a bound-orbit well at $L=4.3$ (for $B=0$ the well disappears at $\alpha\simeq1.13\times10^{-3}$), with orbital energies fixed by the initial radius on the turning-point curve, and remain well within the three-horizon regime discussed in Sec.~\ref{sec:background}, and overlap with the parameter space investigated in \citep{Khan2021R}. Each Poincar\'e section is constructed from 12--60 trajectories initialized along the turning-point curve of Eq.~\eqref{eq:Veff}, providing sufficient resolution of resonance islands and stochastic layers while maintaining a manageable computational cost.

\subsection{Poincar\'e sections}

Because the conformal factor in Eq.~\eqref{eq:Omega} is periodic, we construct
stroboscopic Poincar\'e sections by recording $(r,P_r)$ once per drive
period, at $t=nT_d$ with $T_d=2\pi/\omega_d$ and $n=1,2,\dots$, for each
trajectory in a given panel. Since the integration variable is the proper time $\tau$ while the
drive is periodic in the coordinate time $t$, the section crossings
$t(\tau)=nT_d$ are located by root-finding on the dense output of the
integrator rather than by fixed-step sampling. All frequencies entering
the resonance condition, including $\omega_d$ and the orbital frequency,
are measured with respect to the coordinate time $t$.
The resulting stroboscopic map provides a reduced two-dimensional phase portrait at fixed $(L,\alpha,B,\epsilon,\omega_d)$ from which the
onset of chaos can be read off directly, following standard practice in
the study of order and chaos in both classical and relativistic
Hamiltonian systems \citep{Contopoulos2002}. Sequences of such sections
at increasing $\epsilon$ (or, in turn, $\alpha$ or $B$) constitute the
bifurcation study of Sec.~\ref{sec:results}.

\subsection{Lyapunov exponents and SALI}

To complement the qualitative information provided by the Poincar\'e sections, we quantify the degree of chaos using two independent dynamical indicators. The largest Lyapunov exponent is computed with the standard
two-trajectory renormalization algorithm of Benettin \emph{et al.}~\citep{Benettin1980}: a shadow orbit displaced by $d_0=10^{-8}$ is
integrated alongside the reference orbit, the separation is measured
in the reduced $(r,P_r)$ plane and renormalized every half drive
period, and the exponent is accumulated per unit proper time over
several hundred renormalization segments. As an independent consistency check, the Smaller Alignment Index (SALI) is computed within the same numerical framework, using two
shadow trajectories with initially orthogonal displacements in the
$(r,P_r)$ plane; the finite-difference construction imposes an
indicator floor of order $d_0$, which is ample for the
regular/sticky/chaotic classification employed here.
The SALI \citep{Skokos2001} distinguishes regular orbits, for which SALI decays as a power law in a
two-dimensional map, from chaotic ones, for which it decays
exponentially, on much shorter integration times
than are typically required for the Lyapunov exponent to converge. The
explicit $5\times5$ Jacobian of the right-hand side of
Eq.~\eqref{eq:eom}, needed for both indicators, is given in
Appendix~\ref{app:jacobian}.  Figure~\ref{fig:accuracy} illustrates the long-term preservation of the Hamiltonian constraint for a representative orbit.

\begin{figure}[H]
\centering
\includegraphics[width=0.75\textwidth]{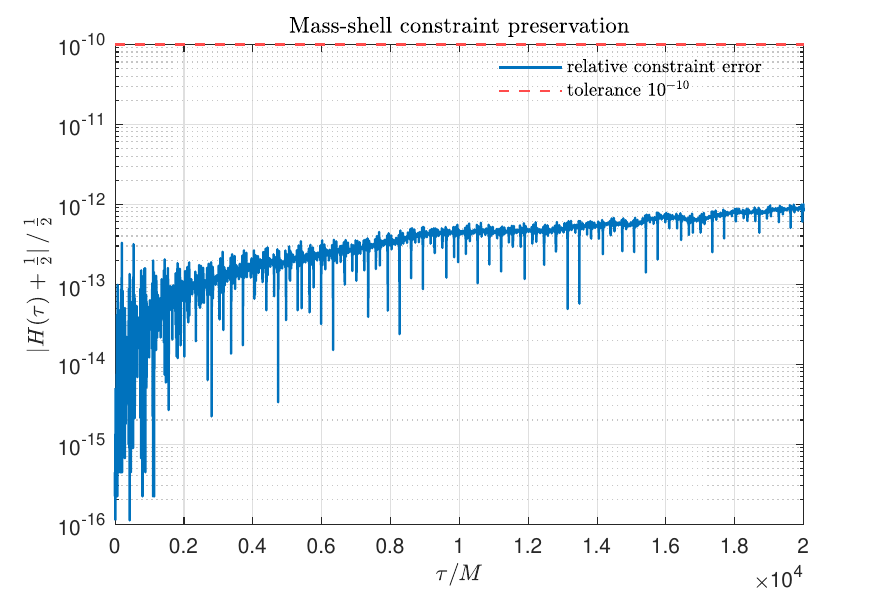}
\caption{\footnotesize \textit{\textbf{Integrator accuracy.} Relative violation
$|H(\tau)+\tfrac12|/\tfrac12$ of the mass-shell constraint along a
representative bound orbit ($r_0=12$, $L=4.3$, $\alpha=0.002$,
$B=0.02$, $Q=0.5$, $\epsilon=0.05$, $\omega_d=0.05$), integrated to
$\tau=2\times10^{4}M$ with the adaptive high-order Runge--Kutta scheme
\textsc{DOP853}. The slow secular growth from $\sim10^{-13}$ to
$\sim10^{-12}$ reflects the non-symplectic character of the scheme,
and remains two orders of magnitude below the acceptance tolerance
$10^{-10}$ (dashed line) over the full window, so the chaos
indicators reported below are not limited by integration error.}}
\label{fig:accuracy}
\end{figure}

\section{Results}
\label{sec:results}

\subsection{Static limit as a consistency check}
\label{sec:static_check}

\begin{figure}[t]
\centering
\includegraphics[width=0.72\textwidth]{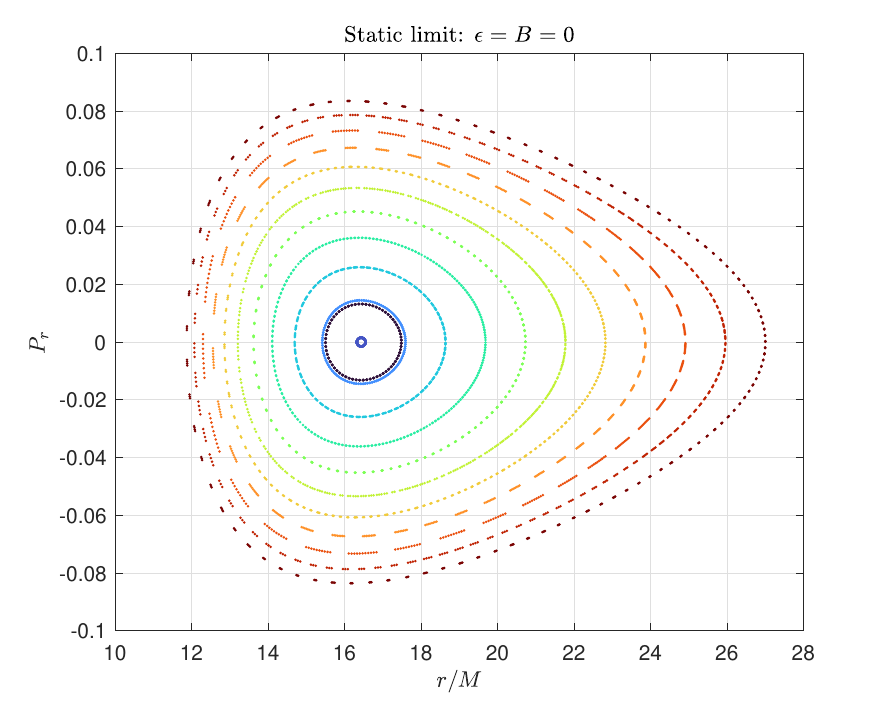}
\caption{\footnotesize \textit{\textbf{Baseline Poincar\'e section, $\epsilon=B=0$.}
Stroboscopic $(r,P_r)$ section for the static quintessential
Reissner--Nordstr\"om case ($\alpha=5\times10^{-4}$, $Q=0.5$, $L=4.3$),
sampled once per reference period $T_d=2\pi/\omega_d$ for twelve
initial radii distributed along the turning-point curve. The section
consists exclusively of smooth nested invariant curves centred on the
stable circular orbit at $r\simeq16.5$, with no island chains or
stochastic layers, confirming the correct reduction of the code to the
integrable limit \citep{Khan2021R} before any time-conformal
or magnetic perturbation is switched on.}}
\label{fig:baseline}
\end{figure}

Before analyzing the periodically driven system, we check that the numerical implementation reproduces the correct dynamics in the appropriate static limit. Formally setting $Q=\alpha=\epsilon=B=0$ recovers geodesic motion in Schwarzschild spacetime, whose Poincar\'e sections consist exclusively of smooth nested invariant tori for bound orbits, without resonant island chains or stochastic layers. Any deviation from this behaviour would therefore indicate a numerical or formulation error rather than a genuine physical effect. In practice, the baseline is run with $Q=0.5$ and a small quintessence parameter, $\alpha=5\times10^{-4}$, so that it connects continuously to the driven runs below. This value is small enough that a bound-orbit well survives at $L=4.3$ with $B=0$. The corresponding Poincar\'e section is presented in Fig.~\ref{fig:baseline}.

\begin{figure}[t]
\centering
\includegraphics[width=\textwidth]{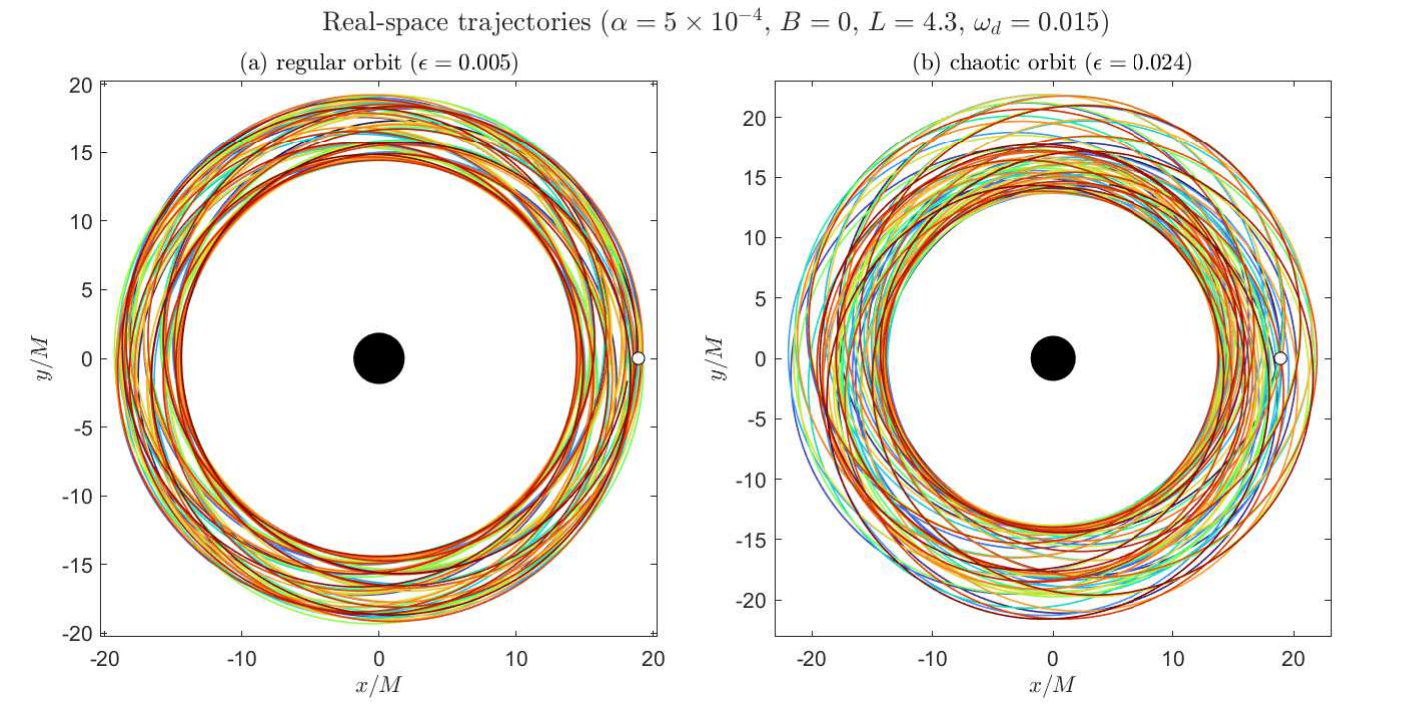}
\caption{\footnotesize \textit{\textbf{Sample real-space trajectories.} Two orbits with
identical launch conditions ($\alpha=5\times10^{-4}$, $B=0$, $L=4.3$,
$\omega_d=0.015$), differing only in the conformal strength, projected
onto the equatorial $(x,y)=(r\cos\phi,r\sin\phi)$ plane and coloured by
proper time; the black disk is the event horizon $r_+$ and the white
marker the common starting point. (a)~Below threshold
($\epsilon=0.005$) the orbit densely fills a sharply bounded annulus,
the real-space image of a regular KAM torus. (b)~Above threshold
($\epsilon=0.024$) the same orbit explores a wider, irregular radial
band with diffuse inner and outer edges, approaching the horizon more
closely, which is the real-space signature of chaotic diffusion in radius. In
an astrophysical setting this radial wandering provides a natural
channel for transporting charged particles between the near-horizon
region and larger radii, of potential relevance to particle
energization in black hole magnetospheres.}}
\label{fig:sample_orbit}
\end{figure}

Retaining the black-hole charge and the quintessence parameter while setting $\Omega\equiv1$ and $B=0$, we recover the static quintessential Reissner--Nordstr\"om metric studied in \citep{Khan2021R}. The largest Lyapunov exponent reproduces, within numerical precision, the stability trends found earlier for bound orbits. Having confirmed agreement with the static quintessential solution, we are now able to add the new ingredients, the time-dependent conformal factor and the magnetic field, and analyze their influence on the chaotic transition.

\subsection{Onset of chaos with conformal perturbation strength $\epsilon$}
\label{sec:eps_results}

Switching on the time-dependent conformal transformation affects the integrable phase space configuration developed for the static case, and leads the system to evolve into chaos. With increasing values of the conformal coefficient $\epsilon$, the results obtained from numerical computations show the expected behaviour dictated by the KAM theorem: the invariant tori are distorted first, then chains of islands develop when the two frequencies are in resonance, and finally neighbouring resonances overlap once $\epsilon$ exceeds a critical value $\epsilon_c$. This is the Chirikov resonance-overlap scenario \citep{Chirikov1979}, and it closely resembles the transition to chaos reported for other nonlinear and relativistic Hamiltonian systems \citep{SunWangLiuWu2021,Dalui2019}.

\begin{figure}[t]
\centering
\includegraphics[width=\textwidth]{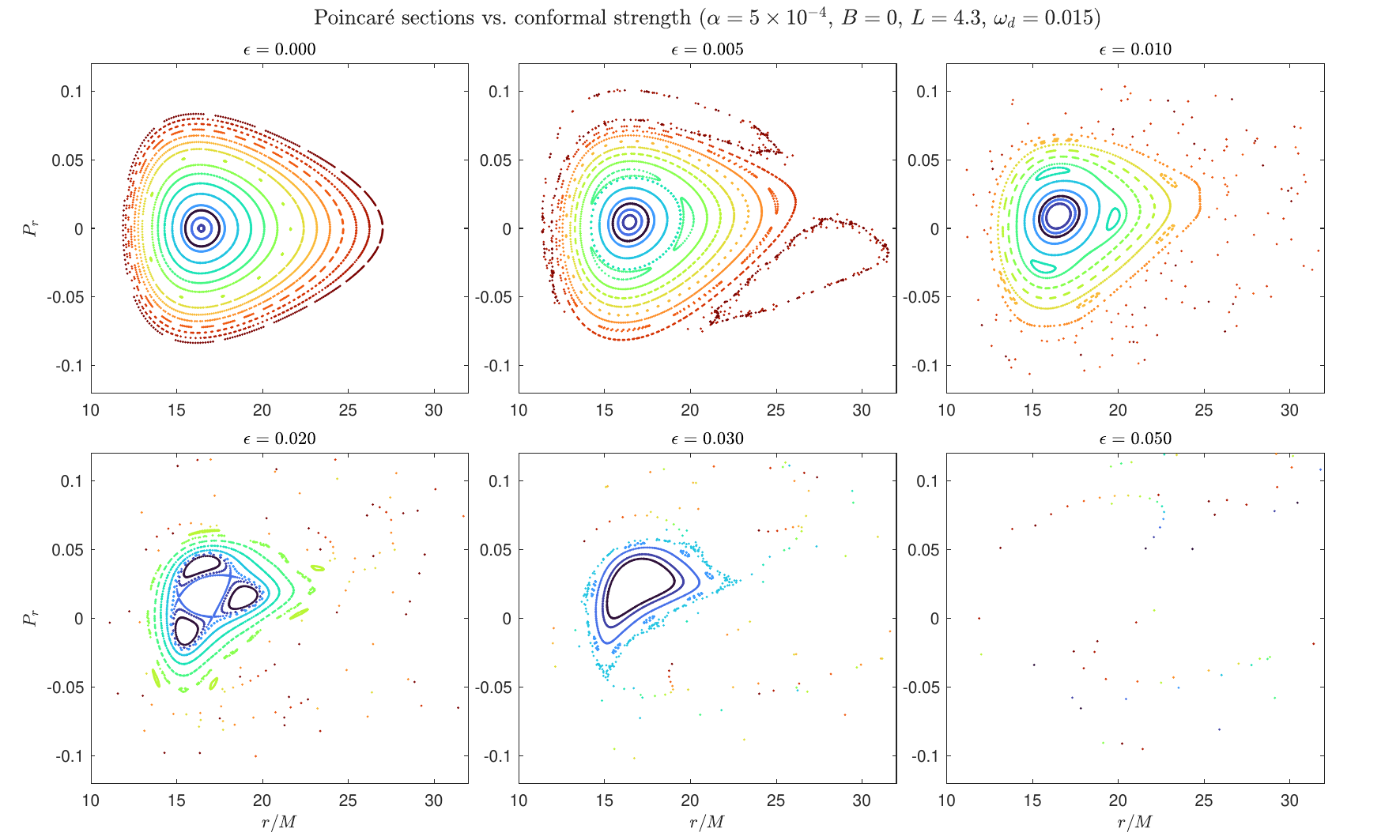}
\caption{\footnotesize \textit{\textbf{Poincar\'e sections vs.\ conformal perturbation
strength $\epsilon$.} Stroboscopic $(r,P_r)$ sections at fixed
$\alpha=5\times10^{-4}$, $B=0$, $L=4.3$, and drive frequency
$\omega_d=0.015$. The panels trace the
standard KAM erosion sequence: nested invariant tori ($\epsilon=0$);
outer-torus break-up and the first resonant island slivers
($\epsilon=0.005$--$0.01$); a fully developed island chain embedded in
a connected stochastic layer ($\epsilon=0.02$); survival of only the
central regular core ($\epsilon=0.03$); and a global stochastic sea
with predominantly escaping orbits ($\epsilon=0.05$). Scattered points
at large $|P_r|$ are the pre-escape footprints of orbits that are
driven over the potential barrier, in accordance with the exact
energy-pumping relation \eqref{eq:Edot}. The innermost core tori are destroyed at
$\epsilon\simeq0.04$, whereas the mid-well threshold
$\epsilon_c\simeq0.023$ entered in Table~\ref{tab:thresholds} is
identified from the bifurcation diagram of
Fig.~\ref{fig:bifurcation_eps}.}}
\label{fig:poincare_eps}
\end{figure}

 While the Poincar\'e sections provide a qualitative visualization of the progressive destruction of the KAM structure, the corresponding bifurcation diagram provides a quantitative measure of the onset of chaos. For the representative mid-well orbit considered here, the transition occurs at $\epsilon_c\simeq0.023$, which serves as a reference threshold for direct comparison with the Lyapunov exponent and SALI diagnostics presented in Sec.~\ref{sec:lyap_results}. The same transition in configuration space is shown in Fig.~\ref{fig:sample_orbit}: above threshold the orbit leaves its sharply bounded annulus and wanders in radius.

\subsection{Role of the quintessence parameter $\alpha$}
\label{sec:alpha_results}

\begin{figure}[t]
\centering
\includegraphics[width=\textwidth]{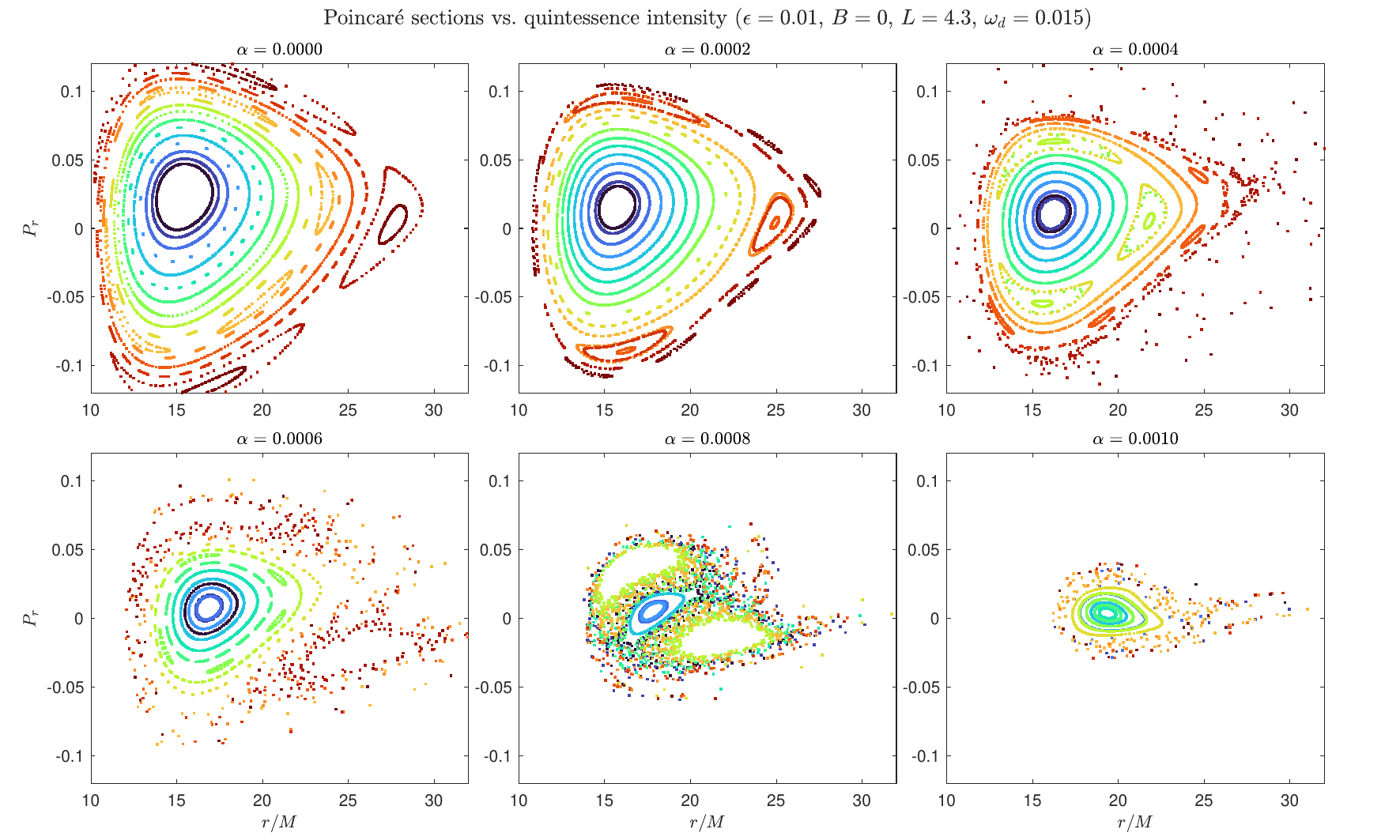}
\caption{\footnotesize \textit{\textbf{Poincar\'e sections vs.\ quintessence intensity
$\alpha$.} Same layout and fixed parameters as
Fig.~\ref{fig:poincare_eps}, now at fixed moderate drive
$\epsilon=0.01$ with $\alpha$ increasing across panels. Growing
quintessence intensity simultaneously thickens the stochastic layer
and shrinks the regular domain, whose centre migrates outward as the
effective-potential well shallows; by $\alpha\simeq10^{-3}$ only a
small regular core survives inside a dissolving chaotic cloud, with
the trailing scatter at large $r$ marking orbits escaping through the
weakened outer barrier. The escape-promoting trend of \citep{Khan2021R} is thus recovered, but it is accompanied, rather than
replaced, by a genuine enhancement of chaos in the surviving bound
population, resolving the question posed in
Sec.~\ref{sec:alpha_results}.}}
\label{fig:poincare_alpha}
\end{figure}

   It was previously shown that increasing the quintessence intensity lowers the effective potential at large radii and therefore facilitates particle escape \citep{Khan2021R}. Whether this escape-promoting behaviour also enhances or suppresses chaos in the periodically driven system is, however, not evident \emph{a priori}. On the one hand, earlier escape may prevent an orbit from exploring phase space long enough to develop clear chaotic signatures; on the other hand, the accompanying modification of the effective potential may strengthen resonance-induced instability. The numerical results presented in Fig.~\ref{fig:poincare_alpha} demonstrate that both effects occur simultaneously: increasing $\alpha$ broadens the stochastic layer of the surviving bound orbits while progressively lowering the escape threshold. As a consequence, for $\alpha \gtrsim 10^{-3}$ the chaotic bound population is increasingly depleted by escape rather than stabilized.

\subsection{Role of the magnetic field $B$}
\label{sec:B_results}

Previous studies of charged-particle motion around magnetized black holes have often identified the magnetic field as a primary source of chaotic dynamics \citep{SunWangLiuWu2021,KopacekKaras2014}. Our results reveal a more nuanced picture for the periodically driven system. At fixed driving frequency, increasing $B$ deepens the effective-potential well and raises the radial frequency. The orbit is first brought into $1{:}1$ resonance with the drive near $B\simeq0.011$, where the chaotic response peaks, and is then detuned from it, so that at stronger fields the stochastic layer contracts and the sections become increasingly regular. Chaos is therefore governed primarily by the resonance condition $n\omega_d\simeq m\Omega_{\mathrm{orb}}(B)$ rather than by the magnetic-field strength alone, with the strongest chaotic response occurring near the intermediate field where the $1{:}1$ resonance is satisfied. This agrees with the small-divisor mechanism discussed in Sec.~\ref{sec:correlation} and indicates that the critical parameter combinations $(\epsilon_c,B_c)$ follow resonance curves in parameter space rather than lines of constant $B$.

\begin{figure}[t]
\centering
\includegraphics[width=\textwidth]{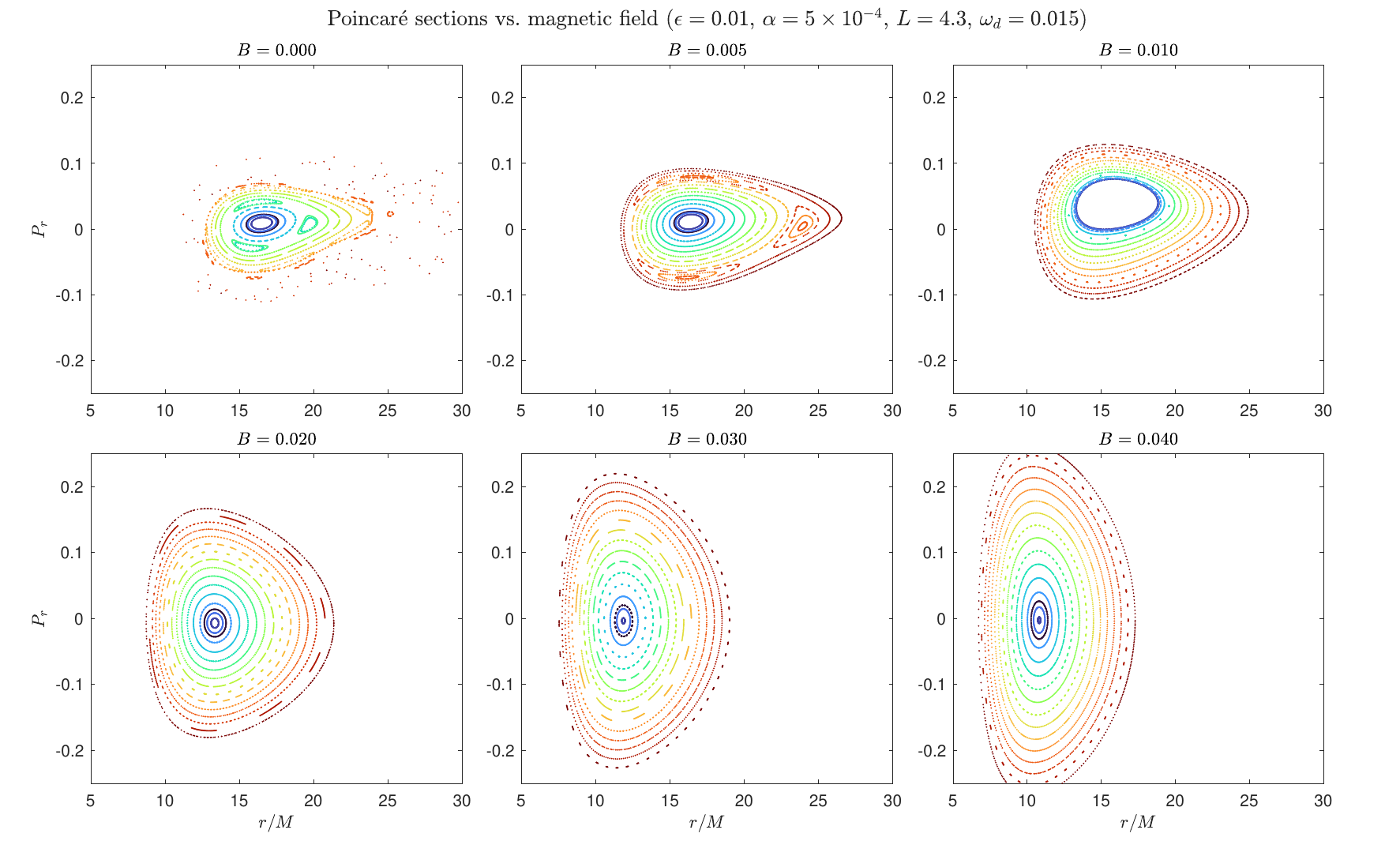}
\caption{\footnotesize \textit{\textbf{Poincar\'e sections vs.\ magnetic field strength
$B$.} Same layout as Fig.~\ref{fig:poincare_eps}, at fixed
$\epsilon=0.01$, $\alpha=5\times10^{-4}$, $L=4.3$, $\omega_d=0.015$,
with initial conditions redistributed around the (inward-migrating)
minimum of the effective potential in each panel. At fixed drive
frequency, increasing $B$ \emph{suppresses} rather than enhances
chaos: the stochastic layer present at $B=0$ shrinks and the sections
become progressively more regular, because the deepening magnetic well
raises the orbital frequency and detunes it from the drive. The large
vacant loop at $B=0.010$ is a $1{:}1$ resonance island traversed by
the innermost trajectories at the intermediate field strength where
the two frequencies momentarily coincide. Chaos is therefore governed
by the resonance condition $n\omega_d\simeq m\Omega_{\mathrm{orb}}$
rather than by the field strength itself, as expected from the
small-divisor mechanism of Sec.~\ref{sec:correlation}.}}
\label{fig:poincare_B}
\end{figure}

\begin{figure}[t]
\centering
\includegraphics[width=\textwidth]{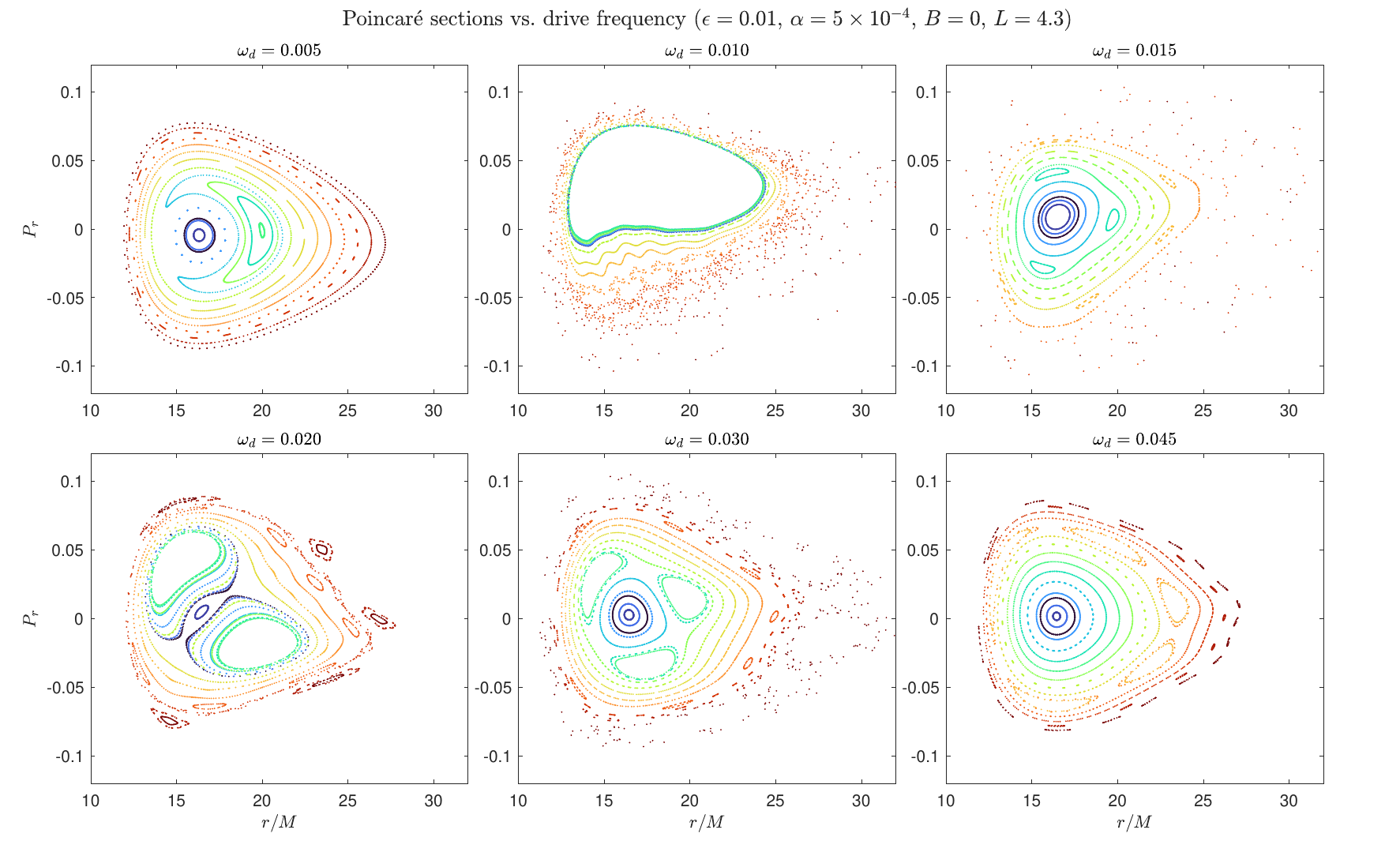}
\caption{\footnotesize \textit{\textbf{Poincar\'e sections vs.\ driving frequency
$\omega_d$.} Fixed $\epsilon=0.01$, $\alpha=5\times10^{-4}$, $B=0$,
$L=4.3$. Chaos is sharply frequency-selective: the stochastic layer is
broadest at $\omega_d=0.010$, which coincides with the radial
small-oscillation frequency $\Omega_{\rm orb}\simeq0.0105$ of the well
centre ($1{:}1$ resonance).
Higher-order responses appear as a two-island chain at
$\omega_d=0.020$ and a three-island chain at $\omega_d=0.030$, while
the sections at both extremes of the scan are essentially regular.
Chaos onset therefore tracks the resonance condition
$n\omega_d\simeq m\Omega_{\rm orb}$ rather than any single parameter
amplitude, consistent with Figs.~\ref{fig:poincare_eps}--\ref{fig:poincare_B}
and with the small-divisor mechanism of Sec.~\ref{sec:correlation}.}
\label{fig:poincare_omega}}
\end{figure}

To confirm that resonance, and not merely the size of the perturbation, drives the transition, we next vary the driving frequency at fixed conformal amplitude. The Poincar\'e sections in Fig.~\ref{fig:poincare_omega} show chaos only inside narrow frequency windows, so it is the matching of frequencies that triggers it.

\begin{figure}[t]
\centering
\includegraphics[width=0.85\textwidth]{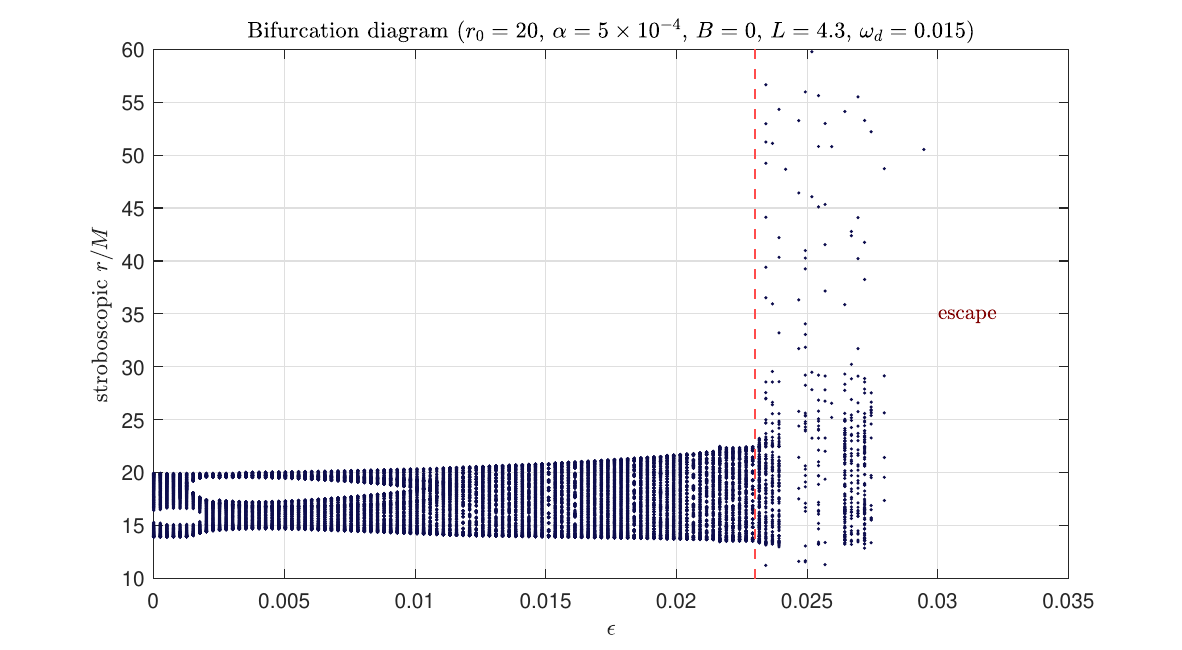}
\caption{\footnotesize \textit{\textbf{Bifurcation diagram vs.\ $\epsilon$.} Stroboscopic
radii of a single mid-well orbit ($r_0=20$, remaining parameters as in
Fig.~\ref{fig:poincare_eps}) recorded once per drive period after
discarding a transient, plotted against the conformal strength. The
orbit remains confined to a thin quasi-periodic band, apart from a brief
resonant band-splitting near $\epsilon\simeq0.002$, up to a sharply
defined threshold $\epsilon_c\simeq0.023$ (dashed line), beyond which
the section points explode into a dense chaotic band connected to the
escape channel; past $\epsilon\simeq0.028$ the orbit escapes within
the sampling window. The threshold is orbit-dependent: this mid-well
value precedes the destruction of the innermost core tori at
$\epsilon\simeq0.04$ visible in Fig.~\ref{fig:poincare_eps}, the two
values bracketing the erosion of the KAM structure from the outside
in.}}
\label{fig:bifurcation_eps}
\end{figure}

\begin{figure}[t]
\centering
\includegraphics[width=\textwidth]{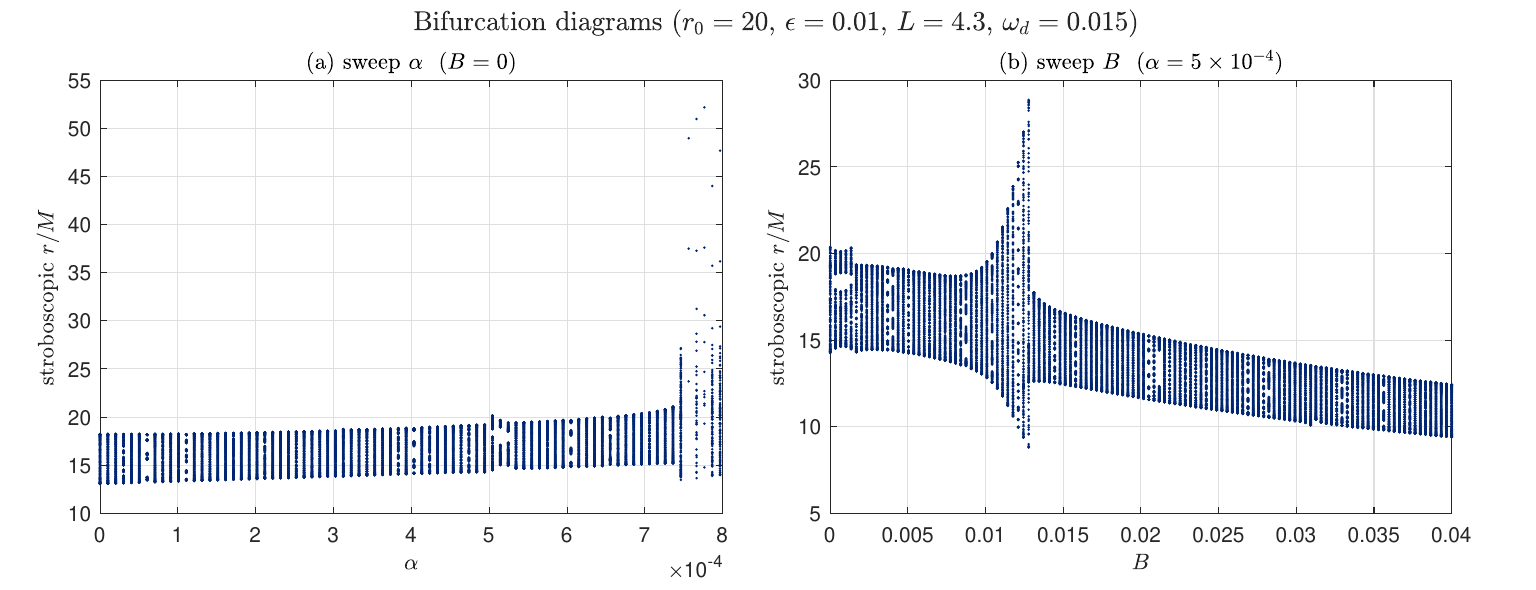}
\caption{\footnotesize \textit{\textbf{Bifurcation diagrams vs.\ $\alpha$ and $B$.}
Stroboscopic radii of a small-amplitude orbit launched at
$r_0=1.15\,r_{\rm min}$ of the instantaneous effective-potential well
for each parameter value, at fixed $\epsilon=0.01$, $L=4.3$,
$\omega_d=0.015$. (a)~Sweeping the quintessence intensity at $B=0$:
the band remains thin up to a sharply defined chaotic burst at
$\alpha_c\simeq7.5\times10^{-4}$, beyond which the orbit escapes through
the eroded outer barrier (empty region); the well itself persists,
increasingly shallow, up to $\alpha\simeq1.13\times10^{-3}$. (b)~Sweeping the magnetic
field at $\alpha=5\times10^{-4}$: chaos is confined to an isolated
resonance burst at $B\simeq0.012$--$0.013$, where the well's rising
orbital frequency crosses the drive; the dynamics is regular on either
side, confirming that $B$ acts through the resonance condition rather
than as a monotonic chaos driver. These thresholds enter the
comparison of Table~\ref{tab:thresholds}.}}
\label{fig:bifurcation_alphaB}
\end{figure}

\subsection{Lyapunov spectrum and SALI maps}
\label{sec:lyap_results}

\begin{figure}[H]
\centering
\includegraphics[width=0.9\textwidth]{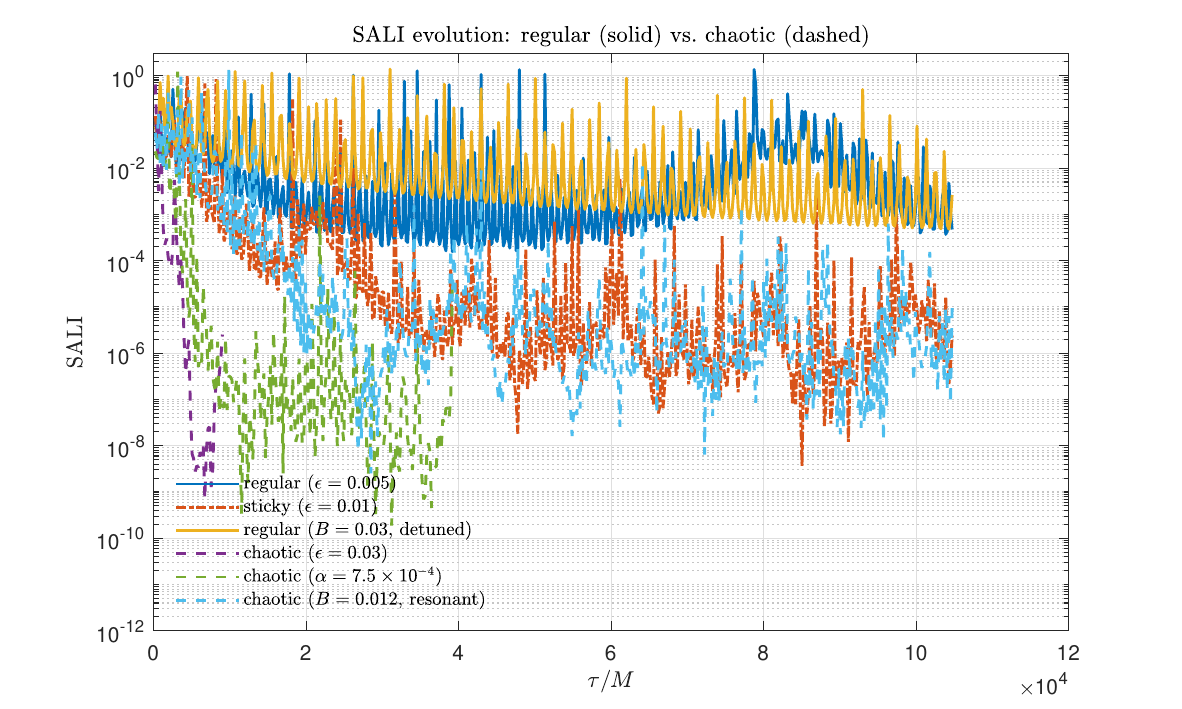}
\caption{\footnotesize \textit{\textbf{SALI time evolution.} Smaller Alignment Index versus
proper time for six representative orbits spanning the parameter
combinations of Figs.~\ref{fig:poincare_eps}--\ref{fig:bifurcation_alphaB},
computed with the same shadow-trajectory scheme as the Lyapunov
estimates (two deviation vectors of magnitude $d_0=10^{-8}$,
renormalized every half drive period). Three dynamical classes
separate cleanly: regular orbits (solid) exhibit the slow power-law
decay with strong oscillations characteristic of quasi-periodic motion
in a two-dimensional stroboscopic map \citep{Skokos2001}; strongly
chaotic orbits (purple, green dashed) collapse exponentially by six to
eight orders of magnitude within $\tau\lesssim2\times10^{4}M$ and
subsequently escape; and near-resonant sticky orbits
($\epsilon=0.01$ mid-well; $B=0.012$ resonant) decay intermediately,
reflecting intermittent trapping near surviving KAM structures. The
floor at $\mathrm{SALI}\sim10^{-8}$--$10^{-10}$ is set by the
finite-difference deviation scheme rather than by machine precision.}}
\label{fig:sali}
\end{figure}

The Poincar\'e sections and bifurcation diagrams presented above characterize the transition to chaos for individual parameter sweeps. To obtain a global view of the dynamics, we now examine two-dimensional maps of the largest Lyapunov exponent and the Smaller Alignment Index (SALI) over the $(\alpha,B)$ and $(\epsilon,\omega_d)$ parameter planes, following a strategy widely adopted in studies of relativistic Hamiltonian chaos \citep{SunWangLiuWu2021,Dalui2019}. These maps also distinguish the well-separated bound-orbit regime considered throughout this work from the asymptotic near-horizon regime, where chaotic behaviour is expected to become increasingly prominent as trajectories approach $r_+$ \citep{Dalui2019}. The time evolution of SALI for representative orbits is shown in Fig.~\ref{fig:sali}, and the resulting classification, summarized in Fig.~\ref{fig:fate_map}, combines the Lyapunov and SALI diagnostics into a unified picture of orbital behaviour across the explored parameter space.

\begin{figure}[t]
\centering
\includegraphics[width=\textwidth]{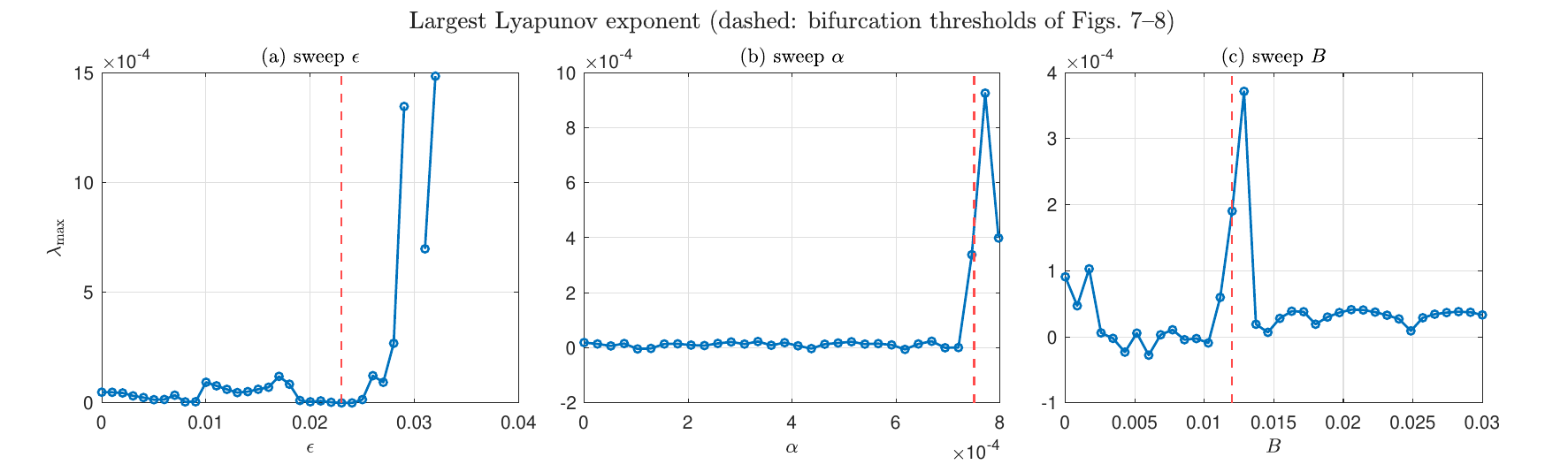}
\caption{\footnotesize \textit{\textbf{Largest Lyapunov exponent vs.\ $\epsilon$, $\alpha$, and $B$.}
Converged Benettin estimates for a small-amplitude orbit launched at
$1.15\,r_{\rm min}$, with the corresponding bifurcation thresholds of
Figs.~\ref{fig:bifurcation_eps} and~\ref{fig:bifurcation_alphaB}
overlaid as dashed lines. In panels (b) and (c), the exponent peaks
at the bifurcation values
$\alpha_c\simeq7.5\times10^{-4}$ and $B_c\simeq0.012$, confirming
quantitative agreement between the two independent diagnostics. In
panel (a), the exponent remains at the noise floor slightly beyond the
bifurcation threshold $\epsilon_c\simeq0.023$ and rises steeply only
at $\epsilon\simeq0.027$. In this narrow interval, the orbit is
chaotic but \emph{sticky}, intermittently trapped near surviving KAM
structures, so that its finite-time Lyapunov exponent lags the
geometric onset of the stochastic band, which is a well-known feature
of near-threshold Hamiltonian chaos \citep{Contopoulos2002}. The
missing point near $\epsilon\simeq0.031$ corresponds to an orbit
escaping within the averaging window.}}
\label{fig:lyapunov_lines}
\end{figure}

\begin{figure}[t]
\centering
\includegraphics[width=0.62\textwidth]{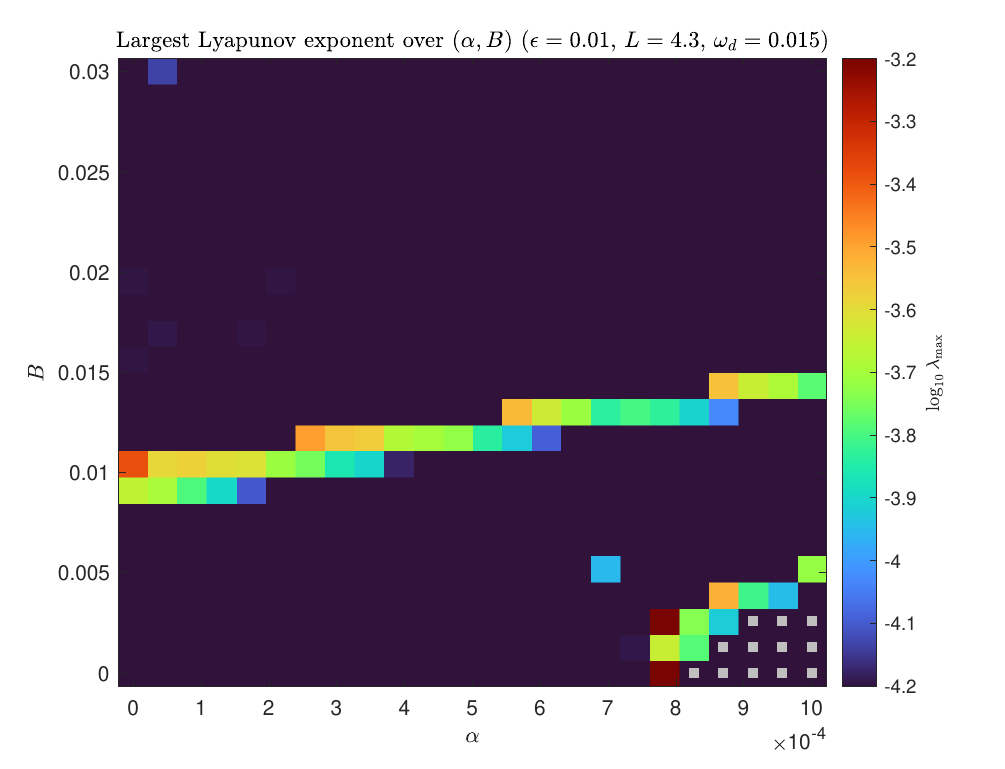}
\caption{\footnotesize \textit{\textbf{Largest Lyapunov exponent over the $(\alpha,B)$
plane} ($\epsilon=0.01$, $L=4.3$, $\omega_d=0.015$). Warm colours mark
chaotic motion, cool colours regular motion; grey cells (lower right)
admit no bound well. The chaotic region forms a narrow ridge that
tilts upward with $\alpha$ (centred at $B\simeq0.010$ for
$\alpha\to0$ and rising to $B\simeq0.014$ at the largest $\alpha$),
tracing the resonance locus $\Omega_{\rm orb}(\alpha,B)=\omega_d$ where
the orbital and drive frequencies coincide. Because increasing
quintessence intensity lowers the orbital frequency, a larger magnetic
field is required to restore resonance, producing the upward tilt. A
separate hot region in the low-$B$, high-$\alpha$ corner marks
barrier-erosion escape rather than resonant chaos, adjacent to the
no-well region. The map confirms that chaos is governed by proximity
to the resonance curve rather than by the magnitude of any single
parameter, in agreement with the tongue structure of
Fig.~\ref{fig:phase_epsomega} and the threshold analysis of
Sec.~\ref{sec:correlation}.}}
\label{fig:phase_alphaB}
\end{figure}

\begin{figure}[t]
\centering
\includegraphics[width=0.6\textwidth]{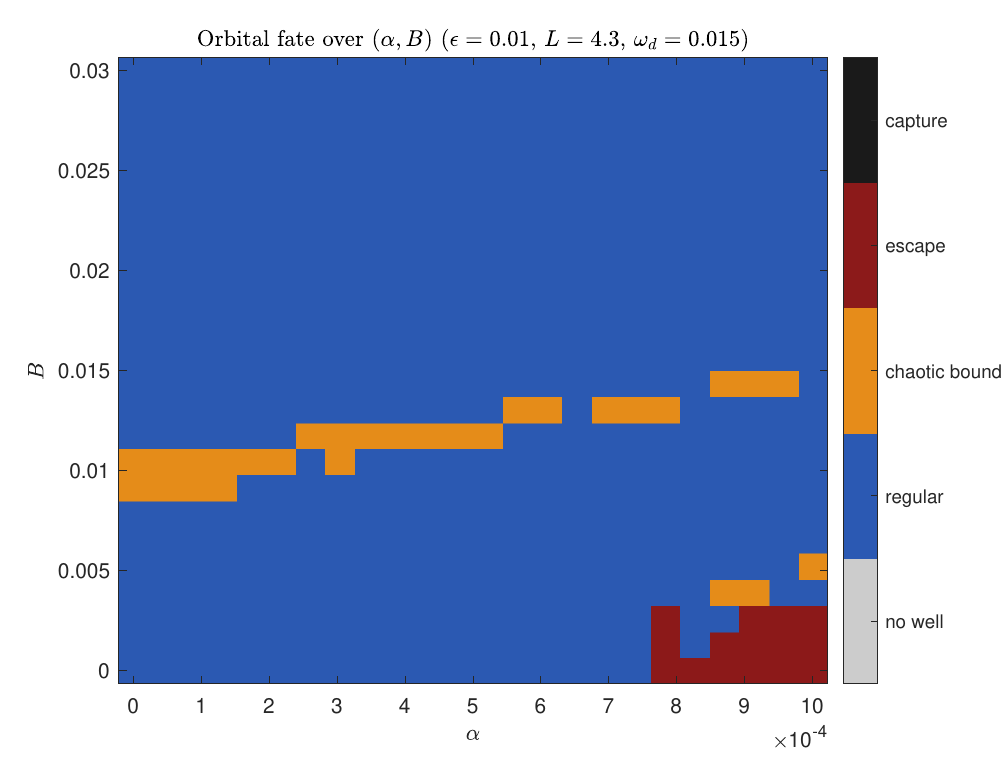}
\caption{\footnotesize \textit{\textbf{Orbital fate classification over the $(\alpha,B)$
plane.} Each cell records the asymptotic fate of an orbit launched at
$1.15\,r_{\rm min}$ of its effective-potential well
($\epsilon=0.01$, $L=4.3$, $\omega_d=0.015$), integrated over $300$
drive periods: regular-bound (blue), chaotic-bound (orange), escaping
to large radius (dark red), or captured by the horizon (black), with
gray marking parameter values admitting no bound well. The
chaotic-bound cells form a diagonal ridge that rises with $\alpha$,
coinciding with the resonance locus $\Omega_{\rm orb}(\alpha,B)=\omega_d$
identified in Figs.~\ref{fig:phase_alphaB}
and~\ref{fig:symmetry_chaos_overlay}, while the escape region occupies
the low-$B$, high-$\alpha$ corner where the quintessence term erodes
the outer barrier. This serves as the parameter-space image of the escape-velocity
trend reported in \citep{Khan2021R}. No horizon capture occurs
for this launch family, so the capture category remains unpopulated;
the grey no-well cells coincide with those of Fig.~\ref{fig:phase_alphaB}.}}
\label{fig:fate_map}
\end{figure}

\begin{figure}[t]
\centering
\includegraphics[width=0.62\textwidth]{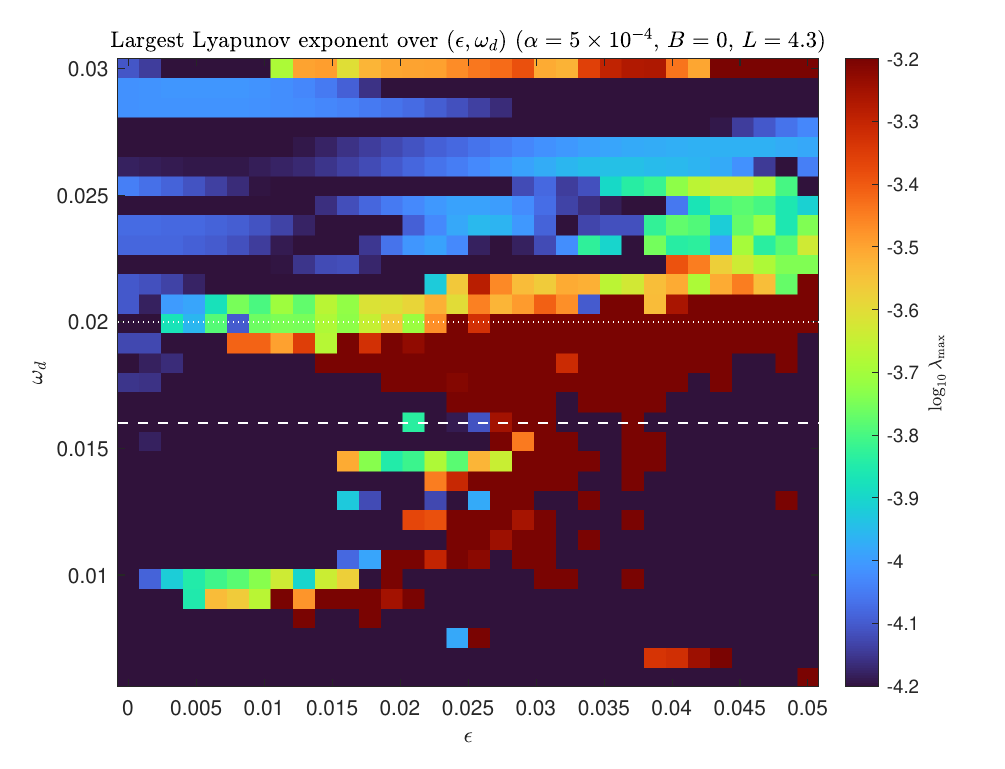}
\caption{\footnotesize \textit{\textbf{Largest Lyapunov exponent over the $(\epsilon,\omega_d)$
plane} ($\alpha=5\times10^{-4}$, $B=0$, $L=4.3$). Chaos (warm colours)
is organized into resonance tongues that emanate from the $\epsilon=0$
axis and broaden with driving strength, as expected for a periodically
driven nonlinear oscillator. The principal tongue (white dashed guide,
$\omega_d\simeq0.0105$) sits at the radial small-oscillation frequency
($1{:}1$), and a secondary tongue (dotted, $\omega_d\simeq0.020$) marks
the $2{:}1$ resonance. Above $\omega_d\simeq0.022$ the motion is regular for all
$\epsilon$ in the scanned range. Such tongues require a periodic drive and
therefore cannot appear in static-background studies of black-hole
chaos.}}
\label{fig:phase_epsomega}
\end{figure}

\subsection{Transient approach to chaos under monotonic $\Omega(t)$}
\label{sec:transient}

\begin{figure}[t]
\centering
\includegraphics[width=\textwidth]{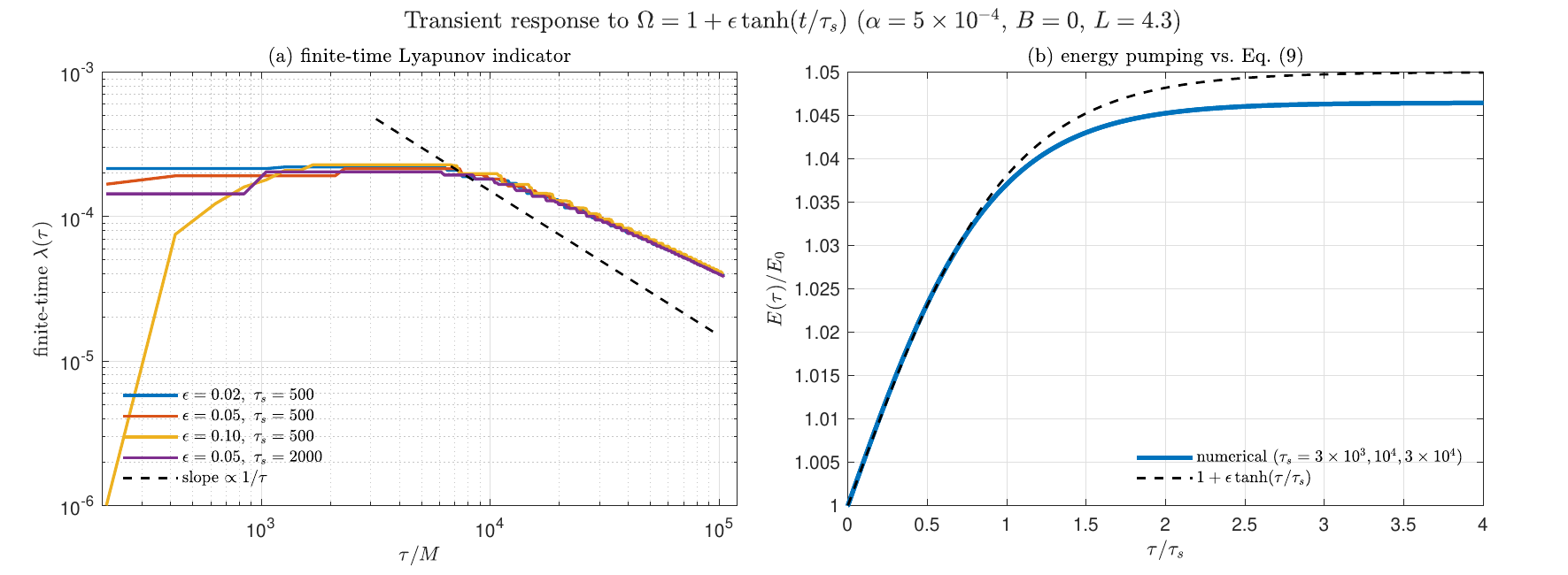}
\caption{\footnotesize \textit{\textbf{Transient response to monotonic conformal driving
$\Omega(t)=1+\epsilon\tanh(t/\tau_s)$} ($\alpha=5\times10^{-4}$, $B=0$,
$L=4.3$). (a)~Envelope of the finite-time Lyapunov indicator: after an
initial switch-on plateau the indicator decays as $\tau^{-1}$ for
every $\epsilon$ and $\tau_s$, so that the orbit settles onto a new
regular torus rather than becoming chaotic. A single adiabatic
deformation, unlike the resonant periodic drive of
Sec.~\ref{sec:eps_results}, cannot sustain the repeated resonant
kicks required to destroy the invariant tori. (b)~Pumped energy
$E(\tau)=-P_t(\tau)$, normalized to its initial value and plotted
against $\tau/\tau_s$: the three $\tau_s$ runs collapse onto a single curve,
confirming that $\tau_s$ sets the energy-transfer timescale. Integrating
Eq.~\eqref{eq:Edot} with $\mathrm{d}\tau=\mathrm{d}t/\dot t$ and
$\dot t=E/(\Omega^{2}f)$ gives a saturated gain
$\Delta E/E_0\simeq f(r)\ln(1+\epsilon)/E_0^{2}\approx0.046$ near the
well centre, which accounts for the plateau at $\simeq0.047$ rather
than $\epsilon=0.05$. This provides the orbital-sector counterpart to the
secular horizon, entropy, and temperature evolution reported in \citep{Khan2022T,Khan2025IJGMMP}.}
\label{fig:transient}}
\end{figure}

To distinguish the effects of resonance from those of time-dependent geometry alone, we finally consider the secular conformal evolution of Eq.~\eqref{eq:Omega_transient} introduced in Sec.~\ref{sec:numerics}. Unlike the periodic modulation studied in the preceding sections, this profile injects energy only during a single smooth transition and therefore does not provide the repeated resonant forcing required for sustained chaotic diffusion. The numerical results shown in Fig.~\ref{fig:transient} confirm this distinction: although the orbital energy evolves on the characteristic timescale $\tau_s$ in agreement with Eq.~\eqref{eq:Edot}, the trajectory ultimately settles onto a new regular orbit rather than becoming chaotic. This behaviour complements the secular evolution of the horizon, entropy, and temperature reported in \citep{Khan2022T,Khan2025IJGMMP}, extending their thermodynamic interpretation to the orbital dynamics of the same spacetime. Time dependence alone is therefore not enough to generate chaos; what is needed is a periodic drive in resonance with the orbit.

\section{Correlating Chaos Onset with Noether Symmetry Breaking}
\label{sec:correlation}

The numerical results of Sec.~\ref{sec:results} establish the parameter ranges over which chaotic motion emerges. We now seek the underlying mechanism by relating these thresholds to the breakdown of the approximate Noether symmetry associated with the energy relation \eqref{eq:Edot}. Following the order-by-order approximate symmetry formalism of Baikov \emph{et al.}~\citep{Baikov1988}, as employed in our companion studies \citep{Khan2022T,Khan2025IJGMMP}, the exact time-translation generator of the static spacetime,
\[
X_0=\partial_t,
\]
is deformed perturbatively according to
\begin{equation}
X = X_0 + \epsilon X_1 + \epsilon^2 X_2 + \cdots,
\label{eq:approx_generator}
\end{equation}
with each $X_n$ determined by requiring that the corresponding correction
$I_n$ to the adiabatic invariant,
\[
\tilde E = E+\epsilon I_1+\epsilon^2 I_2+\cdots,
\]
cancels the $O(\epsilon^n)$ contribution to the energy drift prescribed by
the exact relation $\dot E=\Omega_{,t}/\Omega$ of Eq.~\eqref{eq:Edot}. For the periodic driving of
Eq.~\eqref{eq:Omega}, each correction is obtained through secular
perturbation theory by solving an inhomogeneous equation whose
solvability is governed by the small-divisor factor
$(n\omega_d-m\Omega_{\mathrm{orb}})^{-1}$, where
$\Omega_{\mathrm{orb}}$ denotes the unperturbed radial frequency and
$n,m$ are integers labelling the harmonics of the drive and of the orbit.

When the divisor remains finite, the perturbative correction exists and
the approximate invariant provides an accurate description of the
orbital dynamics. In contrast, when the resonance condition
$n\omega_d\simeq m\Omega_{\mathrm{orb}}$ is satisfied for a low-order
pair $(n,m)$, the correction diverges and the perturbative symmetry
construction breaks down at that radius. This is the same small-divisor
mechanism that underlies the Kolmogorov--Arnold--Moser theorem
\citep{Arnold1963} and the Chirikov resonance-overlap criterion
\citep{Chirikov1979}. Consequently, the order at which the approximate
Noether symmetry first fails is not an independent indicator of chaos,
but the analytical manifestation of the resonance structure responsible
for the destruction of invariant KAM tori observed in
Sec.~\ref{sec:eps_results}.

These considerations motivate the central conjecture tested in this paper:
\emph{the critical perturbation strength $\epsilon_c^{\mathrm{pred}}$ at
which the lowest-order approximate Noether symmetry ceases to admit a
solution coincides, to within the resolution of the numerical scan,
with the critical strength $\epsilon_c^{\mathrm{num}}$ at which the
bifurcation diagram of Sec.~\ref{sec:eps_results} first develops a
chaotic band}, and analogously for $\alpha_c$ and $B_c$. Because
both thresholds trace back to the same small-divisor mechanism, close
agreement would not be a coincidence but a direct consequence of the
shared resonance structure; conversely, a persistent offset between the
two thresholds would indicate that additional, non-resonant routes to
chaos (for instance, overlap of higher harmonics not captured at the
order retained in the approximate-symmetry expansion) are also at
play.

\begin{figure}[H]
\centering
\includegraphics[width=\textwidth]{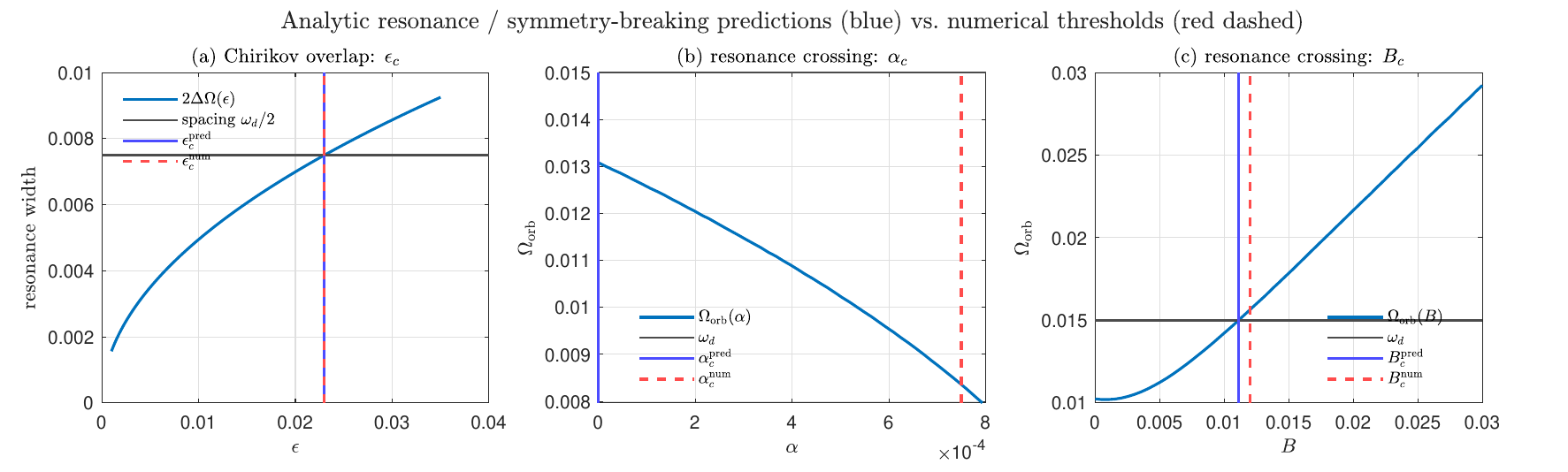}
\caption{\footnotesize \textit{\textbf{Approximate-symmetry / resonance thresholds vs.\
numerical chaos onset.} Blue: analytic predictions from the
small-divisor structure of Sec.~\ref{sec:correlation}; red dashed:
numerical thresholds of Figs.~\ref{fig:bifurcation_eps}--\ref{fig:lyapunov_lines}.
(a)~Chirikov overlap for $\epsilon$: the resonance full width
$2\Delta\Omega(\epsilon)$ reaches the $1{:}1$--$2{:}1$ spacing
$\omega_d/2$ at $\epsilon_c$; the single modulation-coupling constant
is calibrated to $\epsilon_c^{\rm num}$; panel (c) does not use it. (b)~Orbital frequency vs.\ $\alpha$: $\Omega_{\rm orb}$
\emph{decreases} with quintessence intensity and never meets $\omega_d$
in the accessible range, so the $\alpha$-driven transition at
$\alpha_c^{\rm num}\simeq7.5\times10^{-4}$ is \emph{not} a fundamental
resonance but coincides with the erosion of the potential barrier and
the onset of escape, which is the non-resonant route anticipated in
Sec.~\ref{sec:correlation}. (c)~Orbital frequency vs.\ $B$: here
$\Omega_{\rm orb}$ rises through $\omega_d$ at
$B_c^{\rm pred}\simeq0.011$, in close agreement with the measured
$B_c^{\rm num}\simeq0.012$, with no free parameters. The
resonance mechanism therefore reproduces the calibrated $\epsilon$
threshold and predicts the $B$ threshold, while the quintessence threshold is
governed by a distinct barrier-erosion mechanism.}
\label{fig:symmetry_chaos_overlay}}
\end{figure}

To test this conjecture, we compare the critical values predicted by
the approximate symmetry analysis,
$\epsilon_c^{\mathrm{pred}}$,
$\alpha_c^{\mathrm{pred}}$, and
$B_c^{\mathrm{pred}}$,
with the numerical thresholds obtained in
Sec.~\ref{sec:results}. For $\epsilon$, the resonance condition only
locates a resonance, so the threshold is estimated from the resonance
widths, which grow with the perturbation amplitude, together with the
Chirikov overlap criterion \citep{Chirikov1979}. For $B$, the drive
amplitude is fixed and the threshold is set by the field at which the
radial frequency crosses the drive, $\Omega_{\rm orb}(B_c)=\omega_d$. The resulting analytical predictions are compared
directly with the bifurcation, Lyapunov, and SALI diagnostics in
Fig.~\ref{fig:symmetry_chaos_overlay} and summarized in
Table~\ref{tab:thresholds}.

Table~\ref{tab:thresholds} and Fig.~\ref{fig:symmetry_chaos_overlay}
support the proposed link between approximate Noether symmetry
breaking and chaos onset, but the two resonance-driven parameters test
it at different levels. The magnetic threshold is a genuine
prediction: the resonance crossing gives $B_c^{\rm pred}\simeq0.011$
with no adjustable parameter, against $B_c^{\rm num}\simeq0.012$. For
$\epsilon$ the single modulation-coupling constant is fixed from the
numerical threshold itself, so agreement in that row is a consistency
check rather than an independent prediction.

The behaviour of the quintessence parameter is qualitatively different.
Increasing $\alpha$ shifts the orbital frequency away from the driving
frequency rather than towards resonance, so the transition is governed
by the progressive erosion of the outer potential barrier and the
resulting escape of particles instead of resonance overlap. The present
analysis therefore identifies two distinct routes to chaotic behaviour:
a resonance-driven mechanism associated with the breakdown of the
approximate Noether symmetry, and a non-resonant escape mechanism
controlled by the evolution of the effective potential.

\begin{table}[H]
\centering
\caption{\footnotesize \textit{Critical thresholds for chaos onset: analytic
resonance/symmetry-breaking predictions vs.\ numerical measurements.
The $\epsilon$ coupling constant is calibrated to the numerical
$\epsilon_c$; the $B$ prediction is a pure resonance crossing and
uses no fitted constant. The $\alpha$ transition is non-resonant (barrier erosion)
and has no resonance-crossing counterpart.}}
\label{tab:thresholds}
\begin{tabular}{lccl}
\toprule
Parameter & Analytic prediction & Numerical & Mechanism \\
\midrule
$\epsilon_c$ & $0.023$ (calibrated) & $0.023$ & $1{:}1$ Chirikov overlap \\
$B_c$        & $0.011$              & $0.012$ & $\Omega_{\rm orb}(B)=\omega_d$ \\
$\alpha_c$   & -- (no crossing)    & $7.5\times10^{-4}$ & barrier erosion / escape \\
\bottomrule
\end{tabular}
\end{table}

\FloatBarrier

\section{Conclusion}
\label{sec:conclusion}

In this work we investigated the dynamics of a charged test particle in
a periodically time-conformal Reissner--Nordstr\"om--quintessence black
hole spacetime, focusing on a question that has no analogue in a static
background: what determines the transition from regular to chaotic
motion when the spacetime geometry itself evolves in time? Treating the
system as a periodically driven two-degree-of-freedom Hamiltonian
system, we characterized the onset of chaos through several independent
diagnostics, namely Poincar\'e sections, bifurcation diagrams, the
largest Lyapunov exponent, and the Smaller Alignment Index (SALI),
which yield a consistent picture of the transition.
The central result of this work is that the onset of chaos is governed
primarily by resonance rather than by the magnitude of any individual
parameter. Periodic conformal driving progressively destroys the invariant KAM
structure through the familiar sequence of torus deformation, resonance
formation, and stochastic diffusion. The magnetic field modifies this transition indirectly by shifting the orbital frequency relative to the
external drive, so that stronger fields may either enhance or suppress
chaotic behaviour depending on their proximity to resonance. As a
result, the chaotic regions in parameter space organize naturally into
resonance ridges and Arnold-like tongues centred on the condition
$\Omega_{\rm orb}\approx\omega_d$. This resonance-driven organization is
a distinctive feature of periodically evolving black-hole spacetimes and
is absent from the corresponding static models.

The numerical findings are further supported by an analytical
interpretation based on approximate Noether symmetries. The perturbative construction
of the adiabatic invariant is governed by the same small-divisor
structure responsible for resonance overlap, providing a direct
connection between symmetry breaking and the onset of chaotic motion.
The magnetic threshold predicted from the resonance crossing agrees
with the numerical value to within about $10\%$ without any fitted
constant, and the conformal threshold is reproduced once one coupling
constant is fixed. The breakdown of the approximate symmetry therefore
carries physical information about dynamical instability and is not
only a formal property of the perturbation expansion. This extends the role of approximate
Noether symmetries beyond conservation-law analysis to a quantitative
tool for identifying the emergence of chaos in time-dependent black-hole
spacetimes.

Several implications follow. From a dynamical standpoint, the results
suggest that the approximate-symmetry method, developed originally to
assign an energy to time-conformal and gravitational-wave spacetimes,
carries more information than conservation-law accounting alone: the
\emph{failure} of the construction pinpoints where chaos will appear.
From an astrophysical standpoint, the radial wandering of chaotic
orbits and the quintessence-driven escape channel offer a natural
mechanism for transporting charged particles between the near-horizon
region and larger radii, of potential relevance to particle
energization in black-hole magnetospheres and to the slow secular
evolution of orbits around black holes embedded in an accelerating
cosmological background.

A number of directions remain open. The analysis here was confined to
equatorial, weakly magnetized, spherically symmetric motion with the
black hole's own Coulomb field switched off in the particle sector;
relaxing any of these constraints by adding the electrostatic interaction, moving
off the equatorial plane, or extending the analysis to the rotating
Kerr--Newman--quintessence case would provide a stringent test of the
robustness of the resonance and symmetry-breaking correspondence. It would also be
valuable to push the approximate Noether construction to higher order
explicitly, so that the calibrated coupling entering the conformal
threshold could itself be predicted rather than fixed. These extensions will be explored in future work.

\appendix
\section{Variational (tangent-space) Jacobian}
\label{app:jacobian}

The Lyapunov and SALI diagnostics of Sec.~\ref{sec:numerics} require
the evolution of deviation vectors $\delta\mathbf{X}$ in the tangent
space of the flow. Writing the state vector as
$\mathbf{X}=(t,r,\phi,P_t,P_r)^{\mathsf T}$ and the equations of motion
\eqref{eq:eom} as $\dot{\mathbf X}=\mathbf F(\mathbf X,t)$, the
deviation vector obeys the linearized equation
$\dot{\delta\mathbf X}=\mathbf J\,\delta\mathbf X$, where
$\mathbf J=\partial\mathbf F/\partial\mathbf X$ is the $5\times5$
Jacobian evaluated along the reference trajectory. In the numerical
work we integrate deviation vectors by the equivalent two-trajectory
renormalization scheme (Sec.~\ref{sec:numerics}); the explicit
Jacobian below defines the same tangent dynamics and allows direct
reproduction of the variational method.

For compactness we abbreviate $\Omega=\Omega(t)$,
$\dot\Omega=\Omega_{,t}$, $f=f(r)$, $f'=\mathrm{d}f/\mathrm{d}r$,
$f''=\mathrm{d}^2 f/\mathrm{d}r^2$, and introduce the kinetic angular
term
\begin{equation}
\Lambda \equiv P_\phi - qA_\phi = P_\phi - \tfrac{1}{2}qB r^{2},
\qquad
\Lambda' \equiv \frac{\partial\Lambda}{\partial r} = -qB r .
\end{equation}
Because $\phi$ is a cyclic coordinate and $P_\phi$ is conserved, the
flow is independent of $\phi$: the $\phi$ column of the Jacobian
vanishes identically. The $\phi$ row is nonvanishing, but since no
component of the flow depends on $\phi$, the deviation $\delta\phi$ is
slaved to the remaining variables and does not feed back on them.
The evolution equation \eqref{eq:eom_Pphi} decouples, and the
variational dynamics reduces to the four-dimensional subspace
$(t,r,P_t,P_r)$. The non-zero Jacobian elements are listed below.

\medskip
\noindent\emph{Row $\dot t = -P_t/(\Omega^2 f)$:}
\begin{align}
\frac{\partial \dot t}{\partial r}   &= \frac{P_t f'}{\Omega^2 f^{2}}, &
\frac{\partial \dot t}{\partial t}   &= \frac{2 P_t \dot\Omega}{\Omega^{3} f}, &
\frac{\partial \dot t}{\partial P_t} &= -\frac{1}{\Omega^{2} f}.
\end{align}

\noindent\emph{Row $\dot r = f P_r/\Omega^2$:}
\begin{align}
\frac{\partial \dot r}{\partial r}   &= \frac{f' P_r}{\Omega^2}, &
\frac{\partial \dot r}{\partial t}   &= -\frac{2 f P_r \dot\Omega}{\Omega^{3}}, &
\frac{\partial \dot r}{\partial P_r} &= \frac{f}{\Omega^{2}}.
\end{align}

\noindent\emph{Row $\dot P_t = 2H\,\dot\Omega/\Omega$ (off-shell form,
as integrated numerically).} Since the numerical integration employs the off-shell Hamiltonian,
these derivatives remain non-zero away from the constraint surface. Writing $\mathcal K \equiv 2H\Omega^{2}
= -P_t^{2}/f + f P_r^{2} + \Lambda^{2}/r^{2}$, so that
$\dot P_t = \mathcal K\,\dot\Omega/\Omega^{3}$, one finds
\begin{align}
\frac{\partial \dot P_t}{\partial r}
&= \frac{\dot\Omega}{\Omega^{3}}
   \!\left[\frac{f'}{f^{2}}P_t^{2} + f' P_r^{2}
   - \frac{2q A_\phi' \Lambda}{r^{2}}
   - \frac{2\Lambda^{2}}{r^{3}}\right], \\[1mm]
\frac{\partial \dot P_t}{\partial P_t}
&= -\frac{2 P_t\,\dot\Omega}{\Omega^{3} f}, \qquad
\frac{\partial \dot P_t}{\partial P_r}
= \frac{2 f P_r\,\dot\Omega}{\Omega^{3}}, \\[1mm]
\frac{\partial \dot P_t}{\partial t}
&= \mathcal K\!\left(\frac{\ddot\Omega}{\Omega^{3}}
   - \frac{3\dot\Omega^{2}}{\Omega^{4}}\right).
\end{align}
Restricting to the physical constraint surface
$H=-1/2$ recovers the exact relation
$\dot P_t=-\Omega_{,t}/\Omega$ given in
Eq.~\eqref{eq:eom_Pt}. The off-shell form is retained throughout the
variational integration because it provides a consistent propagation of
the tangent dynamics while allowing the mass-shell residual to be
monitored independently (see Sec.~\ref{sec:numerics}).

\noindent\emph{Row
$\dot P_r = -\dfrac{1}{2\Omega^2}\!\left(\dfrac{f'}{f^{2}}P_t^{2}
+ f' P_r^{2}\right)
+ \dfrac{q A_\phi' \Lambda}{\Omega^{2} r^{2}}
+ \dfrac{\Lambda^{2}}{\Omega^{2} r^{3}}$:}
\begin{align}
\frac{\partial \dot P_r}{\partial r}
&= -\frac{1}{2\Omega^{2}}
   \!\left[\left(\frac{f''}{f^{2}}-\frac{2 f'^{2}}{f^{3}}\right)P_t^{2}
   + f'' P_r^{2}\right]
   + \frac{1}{\Omega^{2}}\frac{\partial}{\partial r}
     \!\left(\frac{q A_\phi' \Lambda + \Lambda^{2}/r}{r^{2}}\right), \\[2mm]
\frac{\partial \dot P_r}{\partial P_t}
&= -\frac{f' P_t}{\Omega^{2} f^{2}}, \qquad
\frac{\partial \dot P_r}{\partial P_r}
= -\frac{f' P_r}{\Omega^{2}}, \\[2mm]
\frac{\partial \dot P_r}{\partial t}
&= -\frac{2\dot\Omega}{\Omega}\,\dot P_r .
\end{align}
The remaining $r$-derivative in the first line is evaluated using
$A_\phi'=Br$, $A_\phi''=B$, $\Lambda'=-qBr$, giving
\begin{equation}
\frac{\partial}{\partial r}
  \!\left(\frac{q A_\phi'\Lambda + \Lambda^{2}/r}{r^{2}}\right)
= \frac{q A_\phi''\Lambda + qA_\phi'\Lambda'}{r^{2}}
  + \frac{2\Lambda\Lambda'/r - \Lambda^{2}/r^{2}}{r^{2}}
  - \frac{2\!\left(qA_\phi'\Lambda + \Lambda^{2}/r\right)}{r^{3}} .
\end{equation}

\medskip
Collecting the above derivatives, the Jacobian takes the block form
\begin{equation}
\mathbf J =
\begin{pmatrix}
\partial_t\dot t & \partial_r\dot t & 0 & \partial_{P_t}\dot t & 0 \\[1mm]
\partial_t\dot r & \partial_r\dot r & 0 & 0 & \partial_{P_r}\dot r \\[1mm]
\partial_t\dot\phi & \partial_r\dot\phi & 0 & 0 & 0 \\[1mm]
\partial_t\dot P_t & \partial_r\dot P_t & 0 &
    \partial_{P_t}\dot P_t & \partial_{P_r}\dot P_t \\[1mm]
\partial_t\dot P_r & \partial_r\dot P_r & 0 &
    \partial_{P_t}\dot P_r & \partial_{P_r}\dot P_r
\end{pmatrix},
\end{equation}
where the third column, corresponding to the cyclic coordinate $\phi$,
vanishes identically and the third row decouples from the remaining
dynamics. The resulting $4\times4$
subsystem acting on $(t,r,P_t,P_r)$ is the tangent map that the
two-trajectory scheme of Sec.~\ref{sec:numerics} approximates by finite
differences, and it can be used directly to reproduce the Lyapunov and
SALI results of Sec.~\ref{sec:results}.

\section*{Data Availability Statement}
The datasets generated during and/or analysed during the current study are available from the corresponding author on reasonable request. All theoretical derivations and model parameters are fully contained within the article.

\section*{Declarations}
\paragraph{Conflict of interest}
The authors declare that they have no competing financial or non-financial interests that are directly or indirectly related to the work submitted for publication.

\end{document}